\documentstyle[12pt,psfig]{article}
\begin{document}



\title{ Oxygen in the Earth's core: a first principles study }

\author{ Dario Alf\`e and G. David Price\\
{\sl \small Geological Sciences Dept., University College London,
Gower Street, London WC1E 6BT, U.K.} \\
~~ \\
Michael J. Gillan \\
{\sl \small Physics Department, Keele University, Keele, Staffordshire
ST5 5BG, U.K.} }

\date{\today}
\maketitle

\begin{abstract} 
First principles electronic structure calculations based on density
functional theory have been used to study the thermodynamic,
structural and transport properties of solid solutions and liquid
alloys of iron and oxygen at Earth's core conditions. Aims of the work
are to determine the oxygen concentration needed to account for the
inferred density in the outer core, to probe the stability of the
liquid against phase separation, to interpret the bonding in the
liquid, and to find out whether the viscosity differs significantly
from that of pure liquid iron at the same conditions. It is shown that
the required concentration of oxygen is in the region $25-30$ mol
percent, and evidence is presented for phase stability at these
conditions. The Fe-O bonding is partly ionic, but with a strong
covalent component. The viscosity is lower than that of pure liquid
iron at Earth's core conditions. It is shown that earlier
first-principles calculations indicating very large enthalpies of
formation of solid solutions may need reinterpretation, since the
assumed crystal structures are not the most stable at the oxygen
concentration of interest.

\end{abstract}



\section{Introduction}

The Earth's liquid outer core consists mainly of iron, but its density
is about 10\% too low to be pure iron (Birch, 1952), so that it must
contain some light element. The nature of this element is still
uncertain, and during the last 45 years the main candidates have been
carbon (Birch, 1952; Clark, 1963; Urey, 1960; Wood, 1993), silicon
(Birch, 1952; MacDonald and Knopoff, 1958; Ringwood, 1959, 1961,
1966), magnesium (Alder, 1966), sulphur (Clark, 1963; Urey, 1960;
Birch, 1964; Mason, 1966; Murthy and Hall, 1970; Lewis, 1973), oxygen
(Dubrovskiy and Pan'kov, 1972; Bullen, 1973; Ringwood, 1977), and
hydrogen (Birch, 1952; Fukai and Akimoto, 1983; Suzuki et al., 1989).
For a given light element, it is also uncertain what concentration is
needed to explain the inferred density in the core. The arguments for
and against each of the candidate light elements have been reviewed by
Poirier (1994).

The aim of this paper is to use first-principles calculations to
investigate the possibility that oxygen is the light element.
First-principles calculations are well established as a reliable way
of predicting the thermodynamic, structural and dynamical properties
of solid and liquid materials, including liquid metals (\v{S}tich et
al., 1989; Kresse and Furthm\"uller, 1993). We have recently reported
calculations of this kind on pure liquid iron under core conditions
(Vo\v{c}adlo et al., 1997; de Wijs et al., 1998), which show that it
is a simple close-packed liquid with a viscosity not much greater than
that of many liquid metals at ambient pressure, contrary to some
earlier suggestions (Secco, 1995). We have also used first-principles
simulations to investigate a liquid iron-sulphur alloy under the same
conditions (Alf\`e and Gillan, 1998a). We showed that the properties of
the liquid are scarcely affected by the small sulphur concentration
needed to explain the observed density.

The proposal that oxygen is the light element has a long and
controversial history.  Among the earliest proponents were Dubrovskiy
and Pan'kov (1972) and Bullen (1973), the latter of whom suggested an
outer core composition in the region of Fe$_2$O (equivalent to 12.5
wt$\%$). Ringwood (1977) argued that oxygen should be seriously
considered, and used seismic data to estimate the oxygen content as 28
mol percent (10 wt$\%$). However, it is not completely certain whether
the Fe/O liquid is thermodynamically stable against phase separation
under Earth's core conditions at the composition that would be
necessary to explain the density. It is known that the solubility of
FeO in liquid Fe is very low ($\approx 1$ mol percent) near the
melting temperature of pure iron (1811 K) at atmospheric pressure
(Distin et al., 1971). However, the solubility increases rapidly with
temperature, becoming $6.5 \%$ at 2350 K (Fischer and Schumacher,
1978) and rising to the region of $\approx 35 $ mol $\%$ at 2770 K
(Ohtani and Ringwood, 1984). According to Ohtani and Ringwood (1984),
an extrapolation of the available phase measurements would suggest
that the region of immiscibility disappears entirely above $\approx
3080$ K at atmospheric pressure. It is also well established that the
solubility of FeO in Fe increases with increasing pressure, and that
the partial molar volume of FeO in liquid Fe is lower than that of
pure liquid FeO itself.  Ohtani et al. (1984) used their high pressure
measurements on the solubility of FeO to suggest that the Fe/O system
may show simple eutectic behaviour above a pressure of $\approx 20$
GPa, with no region of liquid immiscibility at any temperature.
Subsequently, Ringwood and Hibberson (1990) showed by direct
measurements that at 16 GPa addition of FeO to pure iron causes a
depression of melting point, leading to a eutectic point at oxygen
mole fraction of 28 $\%$ and a temperature of ca. 1940 K.  Boehler's
(1992) measurements on the melting of Fe/O mixtures are consistent
with these ideas.  Experiments of Knittle and Jeanloz (1991) and
Goarant et al. (1992) on the reaction between lower mantle material
and molten iron at pressures above 70 GPa revealed that the liquid
dissolves significantly amount of FeO.  However, Sherman (1995) has
recently used first-principles calculations on crystalline Fe/O phases
to argue strongly against significant amounts of oxygen in the
core. His calculations gave values for the enthalpy of formation of
crystals of composition Fe$_3$O and Fe$_4$O starting from Fe in the
hexagonal close packed structure and FeO in the NiAs structure.  The
Fe$_3$O and Fe$_4$O compositions were used to model substitutional and
interstitial oxygen respectively. The enthalpies of formation were
found to be so large that phase separation into FeO and Fe appears to
be inevitable. However, it is not clear that Sherman's results have
any relevance to the outer core, since in the liquid phase oxygen does
not have to be either substitutional or interstitial. Even for the
solid phase it is not obvious that Sherman's argument is robust, since
Fe/O crystal structures other than those he studied might well give
much lower enthalpies of formation.

We are mainly concerned in this paper with the Fe/O system
in the liquid state.
Our first-principles calculations, based on density functional
theory and the pseudopotential method, will be used to
address three questions: (a) Is the Fe/O liquid stable against phase
separation under Earth's core conditions? (b) If it is, what
oxygen concentration is needed to reproduce the observed
density? (c) At this concentration, do the structural
and dynamical properties of the liquid differ appreciably
from those of pure liquid iron at the same pressure and
temperature? Our first-principles simulations of the liquid
will provide strong evidence that it is thermodynamically
stable, and that the observed density requires an oxygen
concentration of $25-30$ mol-percent. In studying the properties of
the liquid, we shall be particularly concerned with the
viscosity, since this is one of the most poorly determined
properties of the outer core, with estimates from different
experimental and theoretical methods spanning many orders
of magnitude (Secco, 1995). We shall also investigate a number of other
properties, including the nature of the short-range order,
the atomic diffusion coefficients, and the electronic structure.

Although we are mainly interested in the liquid, we shall also
present some results for the energetics of various crystalline
forms of the Fe/O system. These crystal calculations
serve two purposes: first, they demonstrate that our techniques
are in complete agreement with those used by Sherman in
predicting large enthalpies of formation for
Fe$_3$O and Fe$_4$O in the structures he assumes;
second, they demonstrate that there are crystal structures 
that give much lower formation enthalpies -- an important
fact in understanding how the Fe/O liquid can be stable
against phase separation. It is not our intention to come to
definite conclusions about the possible phase stability of
the Fe/O solid solutions themselves, but our calculations
suggest that their stability cannot be ruled out.

The paper is organised as follows. In section 2, we summarise the
first-principles techniques on which the work is based. Section 3
presents our calculations on the energetics of Fe/O crystals.
In section 4 we report our results on liquid Fe/O, including
its structural properties, the evidence for its stability against
phase separation, and its dynamical and electronic properties.
The final sections present discussion and conclusions.

\section{Methods}\label{method}

In first principles calculations, the solid or liquid is represented
as a collection of ions and electrons, and for any given set of ionic
positions the aim is to determine the total energy and the force on
every ion by solving the Schr\"odinger equation.  This is a formidable
task if the number of atoms is large, but it was made feasible by the
introduction of density functional theory (DFT) many years ago
(Hohenberg and Kohn, 1964; Kohn and Sham, 1965; Jones and Gunnarsson,
1989; Parr and Yang, 1989).  DFT treats electronic exchange and
correlation in a way that allows the electrons to be described by
single-particle wavefunctions, with the interaction between them
accounted for by an effective potential.  DFT can be applied in two
ways: all-electron calculations, or pseudopotential calculations. The
first approach includes such standard techniques as full-potential
linearised augmented plane waves (FLAPW) and linearized muffin-tin
orbitals (LMTO). In the pseudopotential approach, only valence
electrons are explicitly treated, the effect of the core electrons
being included by an effective interaction between the valence
electrons and the cores. In both approaches, the accuracy with which
the real material is described is governed by the approximation used
for the electronic exchange-correlation energy. Until recently, the
local density approximation (LDA) was the standard method. But in
order to achieve the highest accuracy for transition metals it is
essential to use an improved method known as the generalized gradient
approximation (GGA) (Wang and Perdew, 1991). The present work is based
on the pseudopotential approach and the GGA.  A non-technical review
of first-principles calculations based on the pseudopotential approach
has been given recently by one of the authors (Gillan 1997).

There have already been extensive first-principles calculations on
crystalline iron both at ambient pressure and at pressures going up to
Earth's core values (Stixrude et al., 1994; S\"oderlind et al., 1996).
The calculations have been performed using different all-electron
techniques and the pseudopotential technique, and a variety of
properties have been studied, including the equilibrium volume, the
elastic constants, the magnetic moment, the volume as a function of
pressure and lattice vibration frequencies. The agreement between
results obtained with different techniques is generally very close, and
the agreement with experimental data is also good. Particularly
relevant here is the recent comparison of the pseudopotential results
for the pressure-dependent volume of hexagonal-close-packed iron
up to core pressures with earlier all-electron results and with
experimental measurements (Vo\v{c}adlo et al., 1997).

Static first-principles calculations on crystals have been in routine
use for many years. But to study liquids we need to do dynamical
first-principles simulations, in which the calculated forces on the
atoms are used to generate time evolution of the system, with every
atom moving according to Newton's equation of motion.  This kind of
first-principles molecular dynamics (FPMD) pioneered by Car and
Parrinello (1985) has been extensively used to study liquid metals
(\v{S}tich et al., 1989; Kresse and Furthm\"uller, 1993; Holender and
Gillan, 1996; Kirchhoff et al. 1996a, 1996b), and it is known to give
an accurate description of both structure and dynamics.

The present work was performed mainly with the VASP code (Vienna Ab
initio Simulation Package) (Kresse and Furthm\"uller, 1996a,
1996b). As usual in pseudopotential work, the electron orbitals are
represented using a plane-wave basis set, which includes all plane
waves up to a specific energy cut-off.  The electron-ion interaction
is described by ultrasoft Vanderbilt pseudopotentials (Vanderbilt,
1990), which allow one to use a much smaller plane-wave cut-off while
maintaining high accuracy. When we perform FPMD with VASP, the
integration of the classical equation of motion is done using the
Verlet algorithm (1967), and the ground-state search is performed at
each time-step using an efficient iterative matrix diagonalisation
scheme and a Pulay mixer (1980). This method differs from the original
Car-Parrinello technique, which treated the electronic degrees of
freedom as fictitious dynamical variables. In order to improve the
efficiency of the dynamical simulation, the initial electronic charge
density at each time step is extrapolated from the density at previous
steps as described in our previous work (Alf\`e and Gillan, 1998a). In
FPMD on metals, the discontinuity of occupation numbers at the Fermi
level can cause technical difficulties. Since we are interested in
high temperatures in our liquid simulations, the electronic levels are
occupied according to Fermi statistics corresponding to the
temperature of the simulation. This prescription also avoids problems
with level crossing during the ground state search.  The FPMD
simulations are performed at constant temperature (rather than at
constant energy), using the Nos\'e technique (1984).

The iron pseudopotential we use is the same as that used in our
earlier work (Vo\v{c}adlo et al., 1997; de Wijs et al., 1998; Alf\`e
and Gillan, 1998a,b), and was constructed using an Ar core and a
$4s^13d^7$ atomic reference configuration. The oxygen pseudopotential
was constructed using a He core and the $2s^22p^4$ reference
configuration.  At Earth's core pressures the distance between the
atoms becomes so small that the Fe($3p$) orbitals respond
significantly. The net effect is a small repulsion, which we
determined from calculations on the h.c.p. crystal using a Ne core for
the iron pseudopotential instead of Ar. We represent this repulsion by
Fe-Fe and Fe-O pair potentials, as in our earlier work (Vo\v{c}adlo et
al., 1997; de Wijs et al., 1998). The resulting corrections to energy
and forces are generally small.  Non-linear core corrections (Louie et al.,
1982) are included throughout the work.

For all the calculations to be reported, we use a plane-wave cut-off of
400 eV, which gives total energies converged within $\approx 10-20$
meV/atom.  In the calculations on crystals Brillouin-zone sampling is
an important issue, and the sampling density we use is described in
the next section. But for the liquid we use $\Gamma$ point sampling,
which experience suggests should be satisfactory. (We have done
separate tests on pure liquid iron using 4 {\bf k}-points, and we have
found no detectable structural effects, while the average total energy
difference with respect to the $\Gamma$ point only calculations is of
the order of 10 meV/atom, which is completely negligible for the
purposes of the present work).  The time step used in the dynamical
simulations was 1 fs and we generally used a self-consistency
threshold of $1.5 \times 10^{-7}$ eV/atom.  With these prescriptions
the drift of the Nos\'e constant of motion was less than $\approx 60 $
K per ps.

\section{Fe/O solid solutions}\label{solid}

We have used the techniques described in the previous section to
calculate the equilibrium properties and the enthalpy of formation
$\Delta H$ of various members of the Fe/O system, including those
studied by Sherman (1995). Although Sherman's calculations and ours
are both based on DFT, the technical methods are completely different,
since Sherman used the FLAPW method, whereas ours are based on the
pseudopotential approach. One of our main aims is therefore to make
detailed comparisons with his results in order to ensure that the two
methods agree about the energetics of the systems. We shall also point
out that there are Fe$_3$O structures with enthalpies of formation
much lower than those reported previously.

We begin by presenting our results for the equilibrium density
$\rho_0$ (i.e. the density for which the pressure is zero), the bulk
modulus $K$ at this density, and the pressure derivative $K' \equiv
dK/dP$ for crystals of Fe, FeO, Fe$_3$O and Fe$_4$O. Pure Fe is in the
$\epsilon$ structure (hexagonal close packed); FeO is in the B8
structure (the NiAs structure); Fe$_3$O is in the structure obtained
from face-centred cubic Fe by replacing the atoms at the corners of
the conventional cube by O atoms; and Fe$_4$O is in the structure
obtained from f.c.c. Fe by inserting an O atom at the centre of the
conventional cube (see Fig. 1). We have done the
calculations both spin restricted (the occupation numbers of every
electronic orbital are equal for up and down spins, so that there are
no magnetic moments) and spin unrestricted (the occupation numbers for
up and down spins are allowed to vary independently). To sample the
Brillouin zone (BZ) we have used Monkhorst and Pack grids (1976)
using the sampling level that corresponds to 20 and 36 {\bf
k}-points in the irreducible wedge of the BZ respectively for the
cubic (Fe$_3$O and Fe$_4$O) and the hexagonal (Fe($\epsilon$) and
FeO(B8)) crystals.  Using these values, the total energies are converged
within 5 meV/atom for the Fe$_3$O, Fe$_4$O and FeO(B8) structures, and
within $10-15$ meV/atom for the Fe($\epsilon$) structure. We find that,
except for Fe, all the structures are magnetic at low pressures, so
that the system is stabilised if spin restriction is removed. This
disagrees with the results of Sherman (1995), where a significant
magnetic moment was found only for the FeO(B8) structure (a weak
moment was found for Fe$_3$O).

Since we regarded the disagreement concerning magnetic properties as
disturbing, we repeated our calculations of the equilibrium magnetic
moment of Fe$_3$O and Fe$_4$O using a completely independent
electronic structure technique, namely the LMTO method (linearised
muffin-tin orbitals), using the LMTO-46 code due to Krier et
al. (1994). These calculations completely confirm the magnetic
ordering in Fe$_3$O and Fe$_4$O and give numerical values for the
magnetic moments that agree  well with those given by our
pseudopotential calculations. This suggests that the minimisation of
the total energy with respect to magnetic moments may not have been
fully under control in Sherman's work.

For each system we
have calculated the static internal energy $E$ for a series of volumes, and
fitted the results to the Birch-Murnaghan equation of state:
\begin{eqnarray}\label{murna}
E = E_0 + \frac{3}{2}V_0K \left [ \frac{3}{4}(1+2\xi)\left
(\frac{V_0}{V}\right )^{4/3} - \frac{\xi}{2}
\left ( \frac{V_0}{V} \right )^{2} 
 -\frac{2}{3}(1+\xi) \left ( \frac{V_0}{V}
\right )^{2/3} +
\frac{1}{2} \left ( \xi + \frac{3}{2}\right ) \right ] \\
\xi = \frac{3}{4}(4 - K'), \hspace{12cm} \nonumber
\end{eqnarray}
where $K$ is the zero pressure bulk modulus, $K'= (dK/dP)_{P=0}$,
$E_0$ is the equilibrium energy and $V_0$ the equilibrium volume.  Our
calculated values of $\rho_0$, $K$ and $K'$ are compared with
Sherman's results in Table 1, which also reports our
calculated magnetic moments.  The overall conclusion from the
comparison is that the results agree well in the cases where magnetism
is absent: pure Fe and spin-restricted Fe$_3$O and Fe$_4$O (Sherman's
calculations are effectively spin-restricted for Fe$_3$O and Fe$_4$O,
since he found no moments). The equilibrium density agrees in those cases
to better than $2 \%$, and the values of $K$ and $K'$ to about $10
\%$. However, our results show that magnetism has a strong effect on
the equilibrium properties, so that it is important to treat it
correctly. The case of FeO(B8) is problematic. Our spin-unrestricted
calculations give a $\rho_0$ value in respectable agreement with that
of Sherman, but the agreement is poor for $K$ and $K'$. It is clear
that the accurate treatment of the volume dependence of magnetic
moment is important in obtaining reliable values for these parameters,
and our suspicion is that problems with moments may have affected
Sherman's values. However, we shall stress below that magnetic effects
become unimportant at core pressures, so the most important feature of
Table 1 is the good agreement between the two sets of
calculations for the non-magnetic cases.

We turn now to the enthalpies of formation $\Delta H$ of Fe$_3$O and
Fe$_4$O, defined by:
\begin{eqnarray}
\Delta H({\rm Fe_3O}) = H({\rm Fe_3O}) - H({\rm FeO(B8)}) - 2H({\rm
Fe}(\epsilon)), \nonumber \\ 
\Delta H({\rm Fe_4O}) = H({\rm Fe_4O}) - H({\rm FeO(B8)}) -
3H({\rm Fe}(\epsilon)),
\end{eqnarray}
where $H \equiv E + PV$ is the enthalpy per formula unit of each
material. The total energy is taken directly from our Birch-Murnaghan
fit, and the pressure $P = -dE/dV$ is calculated from the derivative
of the fitted form. We have done these calculations both spin
restricted and spin-unrestricted, and our results are reported in the
two panels of Fig. 2.  In the spin-restricted panel we also
show Sherman's enthalpy results. The most important conclusion is that
$\Delta H$ for both Fe$_3$O and Fe$_4$O becomes very large at core
pressures. In the range from 135 GPa (core-mantle boundary) to 330 GPa
(inner-core boundary) $\Delta H$ is at least 3 eV, which corresponds
roughly to a temperature of $3.5 \times 10^4$ K, so that it is
exceedingly unlikely that Fe$_3$O and Fe$_4$O could be
thermodynamically stable in the assumed structures.

Our spin unrestricted results indicate that the true values of $\Delta
H$ are considerably lower than Sherman's results at low pressure, but
at high pressure the differences between the magnetic and non-magnetic
values of $\Delta H$ become very small, so the conclusion is
unaffected. The detailed agreement with Sherman's values of $\Delta H$
is only moderately good, but again this does not affect the
conclusions about the very large size of $\Delta H$.

We now want to ask whether the assumed crystal structures for Fe$_3$O
and Fe$_4$O are actually the most stable. A glance at Wyckoff's book
{\em Crystal Structures} (1964) shows that compounds having the
composition A$_3$B crystallise in a bewildering variety of
structures. We have picked some likely candidates and calculated their
formation enthalpy. Most turn out to be unfavourable, with $\Delta H$
values at least as great as those already reported in
Fig.~2. However, we have discovered one that has a much
lower value. This is the BiI$_3$ structure, which has a rhombohedral
unit cell containing two formula units. Putting Fe$_3$O into this
structure and relaxing both the atomic positions and the shape of the
cell, we end up with the triclinic structure shown in
Fig.~1.  To characterise the structure briefly, we note
that at 300 GPa each oxygen is surrounded by 11 Fe neighbours at
distances of between 1.76 and 2.57 \AA, and two O atoms at distances
of 2.02 and 2.34 \AA~(by contrast, in the cubic Fe$_3$O structure at
the same pressure, each oxygen has 12 Fe neighbours at 2.05 \AA, and
the nearest oxygen neighbours are at 2.9 \AA).  We have calculated the
fully relaxed total energy of this structure at several volumes, and
the resulting structural parameters are reported in Table
2. Spin unrestricted calculations show that the structure
is weakly magnetic, but the moment and the energy stabilisation are so
small that the effects can be ignored.

Fitting of the energies to the Birch-Murnaghan equation of state
yields the enthalpy of formation shown in
Fig. 2. Remarkably, $\Delta H$ is very much lower than for
the previous structures, and it decreases with increasing pressure. At
the pressure of the inner core boundary it is only just over 1
eV. Since we arrived at this distorted BiI$_3$ structure in a rather
haphazard way, it is quite likely that there are other Fe$_3$O
structures with even lower enthalpies. We cannot say at present
whether there are Fe$_3$O structures that are stable against phase
separation under Earth's core conditions, but it certainly does not
look impossible. The low $\Delta H$ for the BiI$_3$ structure will be
highly relevant to our study of phase stability in the liquid Fe/O
system.

\section{The liquid}\label{liquid}

\subsection{Thermodynamics}

Our aim in choosing the thermodynamic parameters for our liquid
simulations was to model liquid Fe/O near the thermodynamic state it
would need to have to reproduce the known outer-core density at the
inner core boundary (ICB).  The temperature at this point is very
uncertain, with estimates ranging from 4000 to 8000 K (Poirier,
1991). We took the value of 6000 K, which is intended to be a
reasonable compromise. However, the density and the pressure are quite
accurately known to be $ \approx 12000$ kg~m$^{-3}$ and $\approx 330 $
GPa.  This density is about $10 \%$ lower than it would be if the core
consisted of pure iron (Birch, 1952).  The main problem in choosing
thermodynamic parameters is that we do not know in advance the
required oxygen concentration, so that a certain amount of trial and
error is needed.

We started from our previous 64-atom simulation for pure liquid iron
(Vo\v{c}adlo et al., 1997), which had a mass density of $13300$
kg~m$^{-3}$ and a calculated pressure of $358 \pm 6$ GPa.  Our first
move was to hold the volume of the system fixed and to transmute the
appropriate number of iron atoms into oxygen atoms to produce the
density of $12000$ kg~m$^{-3}$.  This resulted in a large reduction of
the pressure, and we therefore reduced the cell volume to restore the
original pressure. Naturally, this increased the density, and we
therefore converted more iron atoms into oxygen to regain the density
of $12000$ kg~m$^{-3}$. By repeating this cycle many times, one could
in principle achieve the required density and pressure. But the
calculations are very demanding, since at each state point one has to
equilibrate the system and run it for long enough to obtain adequate
statistics for the pressure, so that in practice a compromise between
accuracy and computational effort is needed.  After several
iterations, we ended up with a simulation box containing 43 iron atoms
and 21 oxygen atoms, i.e. mole fractions of $x_{\rm Fe}\approx 0.67$
and $x_{\rm O}\approx 0.33$.  The resulting mass density of $11600$
kg~m$^{-3}$ and pressure of $342 \pm 4$ GPa are close to the known
values at the ICB.

 Since the mass density of $11600$ kg~m$^{-3}$ is slightly below the
known value at the ICB, it is likely that the concentration of $x_{\rm
O}\approx 0.33$ is an overestimate. We have therefore taken a second
thermodynamic state with a lower concentration. In order to facilitate
comparisons with our calculations on crystalline Fe$_3$O, we chose the
value $x_{\rm O} = 0.25$. This second simulation was performed on a
system of 48 iron atoms and 16 oxygen atoms at the mass density of
$12200$ kg~m$^{-3}$, and the resulting pressure was $366 \pm 8$
GPa. We shall refer to the two simulations in the following as the
`$33 \%$ simulation' and the `$25 \%$ simulation'.

From the thermodynamic results just mentioned, we can estimate the
oxygen concentration that would be needed to reproduce the known
density and pressure at the ICB. Interpolating between the calculated
density values and applying a small correction for the slightly
different pressures in the two simulations, we estimate that the mole
fraction $x_{\rm O} = 0.28$ would reproduce the density $12000$
kg~m$^{-3}$ at the ICB pressure.

In the next sections we describe the structural, dynamical and
electronic-structure properties of the Fe/O liquid alloys.

\subsection{Structure}\label{structure}

We have simulated the $33 \% $ system for 4.2 ps after 2 ps of
equilibration.  The structural properties of the system have been
inspected by looking at the partial radial distribution functions
(rdf), $g_{\rm FeFe}(r)$, $g_{\rm FeO}(r)$, and $g_{\rm OO}(r)$.  The
partial rdf's are defined so that, sitting on an atom of
the species $\alpha$, the probability of finding an atom of the
species $\beta$ in the spherical shell $(r,r+dr)$ is $ 4\pi
r^2n_{\beta} g_{\alpha \beta}(r) dr$, where $n_{\beta}$ is the number
density of the species $\beta$ (the mole fraction of species
$\beta$ times the total number of atoms per unit volume).

We have calculated averages of the rdf's over different small time
windows of the simulation and we find no meaningful differences
between the windows. This confirms that the system is well
equilibrated.  In Fig. 3 we display the rdf's calculated
from the whole simulation. These show that the distance between
neighbouring iron and oxygen atoms is significantly smaller than the
iron-iron distance, the maximum of $g_{\rm FeO}(r)$ being at $\approx
1.7$ \AA, while the maximum of $g_{\rm FeFe}(r)$ is at $\approx 2.1$
\AA.  It is interesting to notice that $g_{\rm OO}(r)$ has a first
maximum at $\approx 2.1$ \AA, which is much greater than the chemical
bond length expected for O-O single or double bonds (1.47 \AA~ and
$1.21$ \AA~ respectively).  This is clear evidence that there is no
covalent bonding between oxygen atoms.  The presence of the O-O peak
at $\approx 2.1$ \AA~ indicates that oxygen atoms repel each other
with an effective atomic diameter of $\approx 2.1$ \AA. This fact shows
that oxygen has two effective sizes in the liquid: a small one when it
interacts with iron and a large one when it interacts with itself.

It is interesting to compare the structural properties of the alloy
with those of pure liquid iron. In Fig. 4 we display
the rdf calculated earlier for pure liquid iron at ICB conditions
(Vo\v{c}adlo et al., 1997) and the $g_{\rm FeFe}$ calculated here.
The two are not very different, the only apparent effect being the
broadening of the peak in the liquid alloy, which is probably due to
the greater disorder in the alloy.

The integration of the first peak of the rdf's provides a definition
of the coordination number $N^c_{\alpha \beta}$ (the average number of
neighbours of species $\beta$ surrounding an atom of species $\alpha$):
\begin{equation}
N^c_{\alpha \beta} = 4 \pi n_\beta \int_0^{r^c_{\alpha \beta}}
r^2 g_{\alpha \beta}(r) dr,
\end{equation}
where $r^c_{\alpha \beta}$ is the position of the minimum after the
first peak of $g_{\alpha \beta}$.  We find the values $N^c_{\rm
FeFe}=11.0$, $N^c_{\rm FeO}=4.5$, $N^c_{\rm OFe}=9.2$, and
$N^c_{\rm OO}=4.5$. For comparison, the average coordination number
found in our earlier simulation of pure liquid iron at ICB conditions
was $N^c_{\rm FeFe}=13.8$ (Vo\v{c}adlo et al., 1997). In interpreting these
numbers, it is helpful to consider the coordination numbers that would
be found if iron and oxygen atoms had exactly the same size and if
atoms were packed in the same way as in pure liquid iron. In that
case, the total number of neighbours of each iron atom, $N^c_{\rm
FeFe} + N^c_{\rm FeO}$, would be the same as in pure iron, whereas in
fact it is 15.5. This increase of coordination number is clearly due
to the smaller size of oxygen, which allows more atoms to be fitted
into the first shell of neighbours. On the other hand, the total
number of neighbours of each oxygen atom, $N^c_{\rm OFe} +
N^c_{\rm OO}$ is 13.7, which is almost the same as the coordination
number in pure iron. We interpret this as the result of two competing
effects. The smaller size of oxygen would lead to a smaller
coordination number if all atoms in its shell of neighbours were
iron. But since on average 4.5 of the neighbours are oxygen, which
have a smaller size when interacting with iron atoms in the shell, the
coordination number is increased again.

We note that the structure of the liquid is very different from that
of the cubic Fe$_3$O and Fe$_4$O crystals discussed in section
\ref{solid}. Oxygen atoms in these crystals have respectively 12 and 6
iron neighbours. The coordination number of 9.2 in the liquid is roughly half way between the two. In the
liquid, the radii from oxygen to iron and oxygen neighbours are equal
to $\approx 1.7$ and $\approx 2.1$ \AA~ respectively, whereas in the
cubic Fe$_3$O crystal at a similar pressure, the distances are 2.05
and 2.9 \AA. On the other hand, in the BiI$_3$-structure Fe$_3$O, the
O-Fe neighbour separation is spread over the range $1.76-2.57$ \AA,
and the O-O separation is in the range $2.02-2.34$ \AA.

We now want to ask whether our simulated system is really in a single
phase and whether we can detect any sign of phase separation. In
studying this it is very helpful to calculate the static structure
factors $S_{\alpha\beta}(k)$ defined by:
\begin{equation}
S_{\alpha\beta}(k) = \langle \rho_\alpha^*({\bf
k})\rho_\beta({\bf k}) \rangle,
\end{equation}
where $\langle \cdot \rangle$ denotes the thermal average (in practice
evaluated as a time average). Here, $\rho_\alpha({\bf k})$ is the
Fourier component of the number density of species $\alpha$ at
wavevector ${\bf k}$, given by:
\begin{equation}
\rho_\alpha({\bf k}) = N_\alpha^{-1/2}\sum_{i=1}^{N_\alpha} {\rm
exp}(i{\bf k} \cdot {\bf r}_{\alpha i}),
\end{equation}
 where $N_\alpha$ is the number of
atoms of species $\alpha$ and ${{\bf r}_{\alpha i}}$ is the position of
the $i$th atom of this species.  Phase separation is associated with
fluctuations of the concentrations of the two species, and the
structure factors give us quantitative information about the
intensities of these fluctuations.

The connection between phase separation and structure factors can be
made more precise. In the limit of zero wavevector, the structure
factors of a liquid alloy can be rigorously expressed in terms of
thermodynamic derivatives (Bhatia and Thornton, 1970):
\begin{equation}
\lim_{ k \rightarrow 0} S_{\alpha\beta}(k) = \frac{k_BT}{(N_\alpha
N_\beta)^{1/2}} \left ( \frac{\partial N_\alpha}{\partial \mu_\beta}
\right )_{V,T,\mu_\beta'}.
\end{equation}
where $\mu_\alpha$ are the chemical potentials, and the notation
indicates that the derivative is to be taken with the volume $V$, the
temperature $T$ and all chemical potential except $\mu_\beta$ held
fixed. But the condition for thermodynamic stability with respect to
phase separation is
\begin{equation}
(\partial \mu_\alpha/\partial x_\beta)_{P,T} > 0.
\end{equation}
At the consolute point (the point in the phase diagram at which phases
start to separate) the derivatives $(\partial \mu_\alpha/\partial
x_\beta)_{P,T}$ become zero.  This implies that the matrix of
derivatives $(\partial \mu_\alpha/\partial N_\beta)_{V,T,N_\beta'}$
has vanishing eigenvalues, corresponding to variations of the numbers
$N_\alpha$ that maintain the pressure constant at fixed volume. But
the matrix $(\partial N_\alpha/\partial \mu_\beta)_{V,T,\mu_\beta'}$ is
the inverse of the matrix $(\partial \mu_\alpha/\partial
N_\beta)_{V,T,N_\beta'}$, so that when the latter becomes singular the
former must acquire infinite eigenvalues. The consequence is that the
values of the structure factors in the zero-wavevector limit must
diverge if the system is unstable with respect to phase
separation. This is also intuitively clear: as one passes from the
miscible to the immiscible region, concentration fluctuations become
ever larger, becoming of macroscopic size when the phases separate,
and the increase in the fluctuations is reflected in the divergence of
the quantities $S_{\alpha\beta}( k \rightarrow 0)$.

Our calculated structure factors for the $33 \%$ simulation are
reported in Fig. 5. They have the form usually found in
liquid alloys, with prominent peaks in $S_{\rm FeFe}(k)$ and
$S_{\rm OO}(k)$ in the region $k \approx 4$ \AA$^{-1}$ signalling the
approximate spatial periodicity associated with the packing of the
atoms. The significant feature for present purposes is the lack of any
anomalous behaviour at small wavevectors. We recognize, of course, that
because of the limited size of the repeating simulation cell, there is
a lower limit to the wavevector that we can examine, which in the
present case is 0.86 \AA$^{-1}$. But at least in the accessible region
of wavevectors there is no indication of any tendency towards phase
separation.

Before leaving the description of structure, we outline another
method we have used to search for signs of phase separation. To
explain this, let us imagine for a moment that the system had
separated into phases of pure Fe and pure FeO. Then the Fe atoms in
the Fe phase would have no oxygen neighbours, whereas the Fe atoms in
the FeO phase might be expected to have 6 oxygen neighbours (we assume
the FeO liquid to have a structure resembling that of crystalline FeO
in the NiAs structure). On the other hand, in the unseparated Fe/O
phase, the number of oxygen neighbours surrounding each Fe atom
fluctuates around the value $4-5$ (see above). We can therefore
distinguish between the two situations by studying the probability
distribution for the number of oxygen neighbours surrounding Fe atoms.
To do this, we use the cut-off distance $r^c_{\alpha \beta}$ defined
above to decide when an atom of species $\beta$ counts as a neighbour
of an atom of species $\alpha$, and we define the function $P_{\alpha
\beta}(n,r^c_{\alpha \beta})$ as the probability that an atom of
species $\alpha$ has $n$ neighbours of species $\beta$. If there is a
complete phase separation, we expect $P_{\rm Fe O}$ to have peaks in the
region of $n=0$ and $n=8$, but if there is no separation we expect a
single peak in the region of $n=4-5$. Note that the rdfs contain less
information than the $P_{\alpha \beta}$ functions, and cannot by
themselves deliver the discrimination we need.

We present in Fig. 6 the function $P_{\rm Fe O}(n,r^c_{\rm
Fe O})$ calculated from our $33 \%$ simulation with the cut-off
$r^c_{\rm Fe O}=2.5$ \AA. This shows a single peak at $n=4$, and no
sign of any structure at lower or higher values of $n$. This means
that there is no indication whatever of any tendency towards phase
separation.

\subsection{Confirmation of phase stability}

Our failure to find any evidence of phase separation strongly suggests
that the Fe/O liquid is in fact stable. But it might be objected that
a simulation lasting only $4-6$ ps does not give enough time for
separation to occur, and that it would occur if the simulation were
longer. To eliminate this possibility we have devised a method in
which phase separation is artificially induced by an external
force. We shall show that when the force is removed the phases
spontaneously re-mix very rapidly. We performed this procedure on the
$25 \%$ simulation (it is not significant that the $25 \%$ case was
chosen for this, and we believe that the $33\%$ system would have
behaved in the same way).

In our procedure, we notionally divide our cubic cell into two parts,
consisting of the left region $0 < x < 0.4$ and the right region $0.4
< x < 1.0$ ($x$ is the coordinate along one of the edge directions in
units of the cell length). The first step is to sweep all the oxygen
atoms into the left region with an external force, so that this region
contains something resembling FeO, while the right region contains
pure Fe. To achieve this, we apply a constant force along the $x$-axis
to all oxygen atoms lying in the right region.  This force is in the
positive $x$ direction for $0.7 < x < 1.0$ and in the negative
direction for $0.4 < x < 0.7$. No force is applied to the iron atoms,
and these are left free to redistribute themselves. Initially, the
magnitude of the force was taken to be 1 eV/\AA, but since this proved
to be too weak it was increased to 3 eV/\AA. After $\approx 1$ ps a
complete phase separation was achieved, with all oxygen atoms in the
left region, and we let the system equilibrate with the external force
still present for a further 1 ps. We show in Fig. 7 a
snapshot of one configuration taken from this period, which clearly
shows the complete separation of phases. The external force was then
switched off and the system was allowed to evolve for a further 2 ps.
Remarkably, we found that after only 1 ps had elapsed the oxygen atoms
became completely randomized throughout the cell, as can be seen in
the snapshot shown in in the lower part of Fig. 7.

To characterise these events quantitatively, we use the probability
function $P_{\rm Fe O}(n,r^c_{\rm Fe O})$ described in section
\ref{structure}. Fig. 8 shows this function calculated by
averaging over three short windows of 0.1 ps each, starting 0 ps , 0.5
ps and 1.0 ps after the external force was switched off. The first
window shows a clear bimodal form, as would be expected for a two
phase system. (We note that because our system is rather small, many
atoms are near the boundary between the phases, so that the peaks in
$P_{\rm Fe O}$ are not as sharp as they would probably be in a larger
system.)  After 0.5 ps, the distribution has already become unimodal,
and after 1 ps it is very similar to what we showed in
Fig. 6 for the equilibrium $33 \%$ simulation. The
conclusion is clear: the system is not stable in the separated state
and returns very quickly to the homogeneous state.
 
In section \ref{solid} we have used the calculations on Fe/O crystals
to show that phase stability in high-temperature Fe/O systems might
well be expected. In particular, we gave an example of an Fe$_3$O
solid structure whose enthalpy of formation is only $\approx 1$ eV. We
have used our $25 \%$ simulation to create another such
structure. This was done simply by quenching the liquid at the rate of
3000 K ps$^{-1}$ until the atoms came to mechanical equilibrium in an
amorphous structure (this is clearly a {\em local} minimum of the
total energy function). We can regard this as a crystal with the
Fe$_3$O composition having an unusually large supercell. The
calculated enthalpy of formation of this solid at the pressure of 290
GPa is reported in the left panel of Fig. 2, and we see
that its stability is even greater that that of the BiI$_3$ structure
proposed for Fe$_3$O in section \ref{solid}. This confirms the idea
that even more stable structures may yet be found.

\subsection{Dynamics}\label{dynamics}

In studying the dynamical properties of the Fe/O liquid, our main
concern is with the viscosity. However, we first give results for the
atomic diffusion coefficients $D_\alpha$, which give a simple way of
characterising the motion of the atoms.
These  are straightforwardly calculated from the mean
square displacement of the atoms through the Einstein relation
(Allen and Tildesley, 1987):
\begin{equation}\label{einstein}
\frac{1}{N_\alpha}\langle \sum_{i=1}^{N_\alpha} |{\bf r}_{\alpha
i}(t_0+t) - {\bf r}_{\alpha i}(t_0)|^2 \rangle \rightarrow 6 D_\alpha
t, ~~ {\rm as} ~~ t \rightarrow \infty,
\end{equation}  
where ${\bf r}_{i \alpha}(t)$ is the position of the $i$th atom of
species $\alpha$ at time $t$, $N_{\alpha}$ has its usual meaning, and
$\langle \cdot \rangle$ is the thermal average, in practice evaluated by
averaging over time origins $t_0$.  In studying the long time
behaviour of the mean square displacement, it is convenient to define
a time dependent diffusion coefficient $D_\alpha(t)$:
\begin{equation}\label{tempo}
D_\alpha(t) = \frac{1}{6 t N_\alpha}\langle \sum_{i=1}^{N_\alpha} |{\bf
r}_{\alpha i}(t_0+t) - {\bf r}_{\alpha i}(t_0)|^2 \rangle,
\end{equation}  
which has the property that
\begin{equation}
\lim_{t\rightarrow \infty} D_\alpha(t) = D_\alpha.
\end{equation}  
In Fig. 9 we display the iron and the oxygen diffusion
coefficients calculated using Eq. (\ref{tempo}). From this data we
estimate $D_{\rm Fe} \approx 0.8 \times 10^{-8}$ m$^2$ s$^{-1}$ and
$D_{\rm O} \approx 1.0 \times 10^{-8}$ m$^2$ s$^{-1}$.  These values
should be compared with those obtained for pure liquid iron, $D_{\rm
Fe} \approx 0.4-0.5 \times 10^{-8}$ m$^2$ s$^{-1}$ (Vo\v{c}adlo et
al., 1997; de Wijs, 1998), and those obtained for Fe and S in our Fe/S
simulation $D_{\rm Fe} \approx 0.4-0.6 \times 10^{-8}$ m$^2$ s$^{-1}$,
and $D_{\rm S} \approx 0.4-0.6 \times 10^{-8}$ m$^2$ s$^{-1}$ (Alf\`e
and Gillan, 1998a).  This means that the two species of atoms in liquid
Fe/O diffuse somewhat more rapidly than the atoms in both liquid iron
and the liquid Fe/S alloy at the same pressure and temperature.

In the past, atomic diffusion coefficients have often been used to
estimate the viscosity of liquids via the Stokes-Einstein relation,
and this was the procedure used in our previous work on liquid Fe and
Fe/S. Since the diffusion coefficients are larger in the present case,
this procedure would lead us to expect a lower viscosity for liquid
Fe/O. We have recently demonstrated (Alf\`e and Gillan, 1998b) that the
viscosity can be more directly (and more rigorously) calculated in
first-principles simulations using the Green-Kubo relations, i.e. the
relations between transport coefficients and correlation functions
involving fluxes of conserved quantities (Allen and Tildesley, 1987).
The shear viscosity $\eta$ is given by:
\begin{equation}\label{shear}
\eta = \frac{V}{k_BT} \int_0^\infty dt \langle P_{xy}(t_0+t) P_{xy}(t_0)
\rangle,
\end{equation}
where $V$ is the volume of the system and $P_{xy}$ is the off-diagonal
component of the stress tensor $P_{\alpha\beta}$ ($\alpha$ and $\beta$
are Cartesian components).  The stress tensor is straightforwardly
calculated, so that the stress autocorrelation function (SACF) $\langle
P_{xy}(t_0+t) P_{xy}(t_0) \rangle$ can also be obtained, but at first
sight it might appear that very long simulations would be needed to
gather adequate statistical sampling. However, we have recently shown
that perfectly adequate $\eta$ values can be obtained with
surprisingly short runs, and we have reported results for liquid
aluminium and liquid Fe/S (Alf\`e and Gillan, 1998b).

In the left panel of Fig. 10 we display the the average of the five
independent components of the traceless SACF divided by its value at
$t=0$ which we denote by $\phi(t)$.  Since the traceless part of the
stress tensor has zero average, $\phi(t)$ goes to zero as $t
\rightarrow \infty$.  The statistical error on $\phi(t)$ for all
values of $t$ is $\approx 5\%$ of the value at $t=0$, and after $
0.2-0.3$ ps the magnitude of $\phi(t)$ falls below that error.  In the
right panel of Fig. 10 we display the integral $\int_0^t dt'\phi(t')$
of $\phi(t)$ as a function of time. The limiting value of the integral
for $t\rightarrow \infty$ is the shear viscosity. The error that one
makes in evaluating that integral grows with time, since one
integrates the noise together with $\phi(t)$.  We estimated the error
in the integral as a function of time using the scatter of the
SACF's. Combining this estimate with an analytic expression for the
error, we obtain the error estimate displayed in Fig. 10.  From the
point where $\phi(t)$ falls below the noise one integrates only the
latter, so one gains nothing by evaluating the integral beyond that
point.  If we assume that $\phi(t)$ is zero above $t\approx 0.3$ ps,
we obtain the value $\eta = 4.5 \pm 1.0 $ mPa~s.  This value is
roughly half the value reported earlier for liquid Fe (Vo\v{c}adlo et
al., 1997) and Fe/S (Alf\`e and Gillan, 1998a) at ICB conditions, and
is not very much greater than the viscosity of typical liquid metals
under ambient conditions; for example, the viscosity of liquid
aluminium at atmospheric pressure 100 K above its melting point is
1.25 mPa~s (Shimoji and Itami, 1986).

This result confirms our earlier conclusion (de Wijs et al., 1998) that the
viscosity of the outer core is towards the lower end of the wide range
of theoretical and experimental values reported in the literature.

\subsection{Electronic structure}\label{electronic}

We have studied the electronic structure of our simulated Fe/O liquid
in order to shed light on the nature of the bonding between the atoms
and to help to interpret the structure discussed in section
\ref{structure}. The main tools used in this analysis are the
electronic density of states (DOS) and the local density of states
(LDOS). The DOS represents the number of electronic states per unit
energy as a function of energy, while the LDOS is a projection of the
DOS onto states of chosen angular momentum on atoms of chosen
species. In performing this projection we took spherical regions of
radius $R$ on the atoms, and in practice we chose $R=0.6$ \AA~ for
both iron and oxygen. This $R$ value is considerably smaller than half
the interatomic distance in the liquid (see Fig. 3), so we
expect to distinguish clearly between the electronic structures on
different atoms. The results are not averaged over time but are
calculated from the electronic energies and wavefunctions for selected
time steps taken from the $25 \%$ simulation.

The calculated DOS is shown in the upper panel of Fig. 11,
and consists of four main features (energies are referred to the Fermi
energy): two fairly narrow peaks at $-24$ and $-11$ eV; a large
dominant peak spanning the range $-9$ eV to $3$ eV; and a broad
feature extending well above the Fermi energy. The LDOS shown in the
lower panel of the Figure allows us to interpret these features. The
peak at $-24$ eV consists entirely of O$(2s)$ states. The peak at
$-11$ eV is mainly O$(2p)$, but with an appreciable contribution of
Fe$(3d)$. The dominant peak from $-9$ eV to $3$ eV consists mainly of
Fe$(3d)$ states, but there is a significant peak associated with
O$(2p)$ states just above the Fermi energy. The broad feature above
the Fermi energy comes from both Fe$(3d)$ and O$(2p)$ states.

To see the implications of this structure for the interatomic bonding,
we note that if the O$(2p)$ states were much lower in energy than the
Fe$(3d)$ states, there would be very little hybridisation between the
two kind of states, the O$(2p)$ levels would be completely filled, and
the oxygen atoms would carry a net charge of $-2|e|$. The Fe-O bonding
would be ionic. On the other hand, if the O$(2p)$ states were at the
same energy as Fe$(3d)$, we should expect strong hybridisation and
little charge transfer, so that the bonding would be covalent. It is
clear from Fig. 11 that the Fe-O bonding is intermediate
between ionic and covalent. The O$(2p)$ states are below Fe$(3d)$, but
not low enough to suppress hybridisation. There is a clear splitting
of the O$(2p)$ levels into bonding and anti-bonding orbitals, though
the bonding orbital clearly has much larger weight (in pure covalent
bonding we should expect the weights to be equal). The peak in
Fe$(3d)$ at $-11$ eV is also clear evidence for hybridisation between
O$(2p)$ and Fe$(3d)$. The implication is that there is a partial
charge transfer from Fe to O, but not enough to give oxygen a charge
of $-2|e|$.

The bonding between Fe atoms is metallic. The partial filling of the
Fe$(3d)$ levels gives the well known bonding mechanism emphasised by
Friedel's (1969) analysis of the cohesive and elastic
properties of transition metal crystals.

Since the oxygen atoms carry partial charges and their $2p$ orbitals
are almost full, no covalent bond is expected between them, and this
is also clear from Fig. 3. To investigate the electronic
state of oxygen in more detail, we have calculated the LDOS for oxygen
atoms in different environments. We show in Fig. 12 the LDOS
for two oxygen atoms denoted O$_{\rm a}$ and O$_{\rm b}$, which have
been chosen from the $25 \%$ simulation so that O$_{\rm a}$ has 10 Fe
neighbours and one O neighbour while O$_{\rm b}$ has seven Fe and four
O neighbours. If there were any covalent bonding between O atoms, we
should expect a larger bonding-antibonding splitting of the $2p$
states for the O$_{\rm b}$ atoms than for O$_{\rm a}$, and we might
expect a similar splitting (or at least a broadening) of the O$(2s)$
states for O$_{\rm b}$. The LDOS curves show neither of these
effects. Instead, the main difference is the upward shift of the peaks
for O$_{\rm b}$ compared with O$_{\rm a}$. We believe this is direct
evidence for the partial charge transfer: the valence electrons of
O$_{\rm b}$ feel a repulsive electrostatic potential due to the
partial negative charges on the four oxygen neighbours, which raises
their energy.

The main bonding mechanisms in the liquid are therefore the ionic-covalent
Fe-O bond and the metallic Fe-Fe bond. Since the O$(2p)$ atomic
orbitals are more compact than Fe$(3d)$ orbitals we expect the Fe-O
distance to be shorter than the Fe-Fe distance, and this effect is
clear from the rdf's shown in Fig. 3. The partial charge
transfer presumably also contributes to the shortening of this
distance.

\section{Discussion}

Two of our main aims in this paper have been to probe the phase
stability of liquid Fe/O under Earth's core conditions, and to
determine the oxygen concentration that would be needed to reproduce
the known core density. In practice, these aims must be taken
together: we want to know whether the alloy is thermodynamically
stable at the appropriate concentration.

The results we have presented leave little doubt that {\em if} the
liquid is stable then the mole fraction of oxygen must be in the
region $25-30 \%$ (our best estimate is $28 \%$), because anything
much less than this would give a pressure that is too low at the known
density.  This is essentially the same as the value proposed many
years ago by Ringwood (1977).  In judging the robustness of this
conclusion we recall some important facts: First, DFT
electronic-structure methods of the type used here generally predict
the density of materials at a given pressure to within a few
percent. Particularly relevant here are recent DFT calculations
(Stixrude et al., 1994; S\"oderlind et al., 1996; Vo\v{c}adlo et al.,
1997) on h.c.p. iron over the pressure range $0-350$ GPa which are in
excellent agreement with each other and with the experimental
results. Similar comparisons for FeSi (Vo\v{c}adlo et al., 1997) are
also relevant. DFT calculations on oxides (including transition-metal
oxides) generally predict the density with similar accuracy. A second
important fact is that our earlier first principles simulations of
pure liquid iron, based on exactly the same techniques, gave a
prediction of the density at the pressure of the inner core boundary
which is correct to $\approx 2 \%$. (In fact the comparison was done
the other way round: at the density of 13300 kg~m$^{-3}$ and the
temperature of 6000 K, our simulations gave a pressure of 358 GPa,
compared with the value of 330 GPa estimated from experimental data.)
Third, our estimate of the oxygen concentration is based on {\em
changes} of pressure and density compared with our simulated pure
iron. The calculations should be even more reliable for these
differences then they are for the absolute values. We therefore
believe that our value for the oxygen concentration required should be
subject to an error of no more that $\approx 5 \%$.

It is interesting to compare the liquid composition with that for the
case of Fe/S. In our recent simulations (Alf\`e and Gillan, 1998a), we
showed that liquid Fe/S is not far from reproducing the known pressure
and density at the inner core boundary with a sulphur mole fraction of
$18 \%$. At this composition, the mass density of 12330 kg~m$^{-3}$
gave a calculated pressure of $349 \pm 6$ GPa.  If we make a
correction to bring the density to 12000 kg~m$^{-3}$, we find a
sulphur mole fraction of $23 \%$. This means that to achieve the
required reduction in density we actually need a higher mole fraction
of oxygen than of sulphur, even though the atomic mass of oxygen is
only half that of sulphur. The reason is that the oxygen atom is
smaller, a point to which we return below.

The question of phase stability is more complex. What seems certain is
that earlier arguments against phase stability based on {\em ab
initio} calculations of the energetics of Fe$_3$O and Fe$_4$O crystals
(Sherman, 1995) do not really deliver the intended conclusion. This is
not because the calculations were wrong. On the contrary, our
calculations fully support their correctness. It is simply that the
crystal structures assumed were not the most stable. We have presented
an alternative structure for Fe$_3$O which gives a formation enthalpy
that is low enough ($\approx 1$ eV) to make phase stability under core
conditions quite plausible.

We have used our simulations to probe the phase stability of the Fe/O
liquid in the appropriate region of concentration, and all the
indications are that it is stable.  We therefore fully support the
conclusion that has been drawn from high pressure experimental
measurements (Ohtani et al., 1984; Ringwood and Hibberson, 1990) that
liquid Fe and FeO are miscible under core conditions in the
concentration region of interest.  However, a word of caution is in
order. The fact is that our simulated systems are rather small, and it
is quite conceivable that a system that would be unstable in the
thermodynamic limit could be stabilised by the artificial periodic
boundary conditions in small simulation cells. Nevertheless, our
calculations certainly provide strong support for the thermodynamic
stability of liquid Fe/O under core conditions at the relevant
concentration. In the end, we believe that the definitive theoretical
approach to this question will be first principles free energy
calculations on the appropriate solid and liquid phases. These would
be computationally very demanding, but should certainly be feasible in
the near future. (First principles free energy calculations have
recently been used with success to calculate the melting properties of
silicon and aluminium (Sugino and Car, 1995; de Wijs et al., 1998).)

The small size of oxygen is clear from our analysis of the liquid
structure: the Fe-O nearest neighbour distance (the position of the
first peak in the rdf) is only at $\approx 1.7$ \AA, compared with
$\approx 2.1$ \AA~ for the Fe-Fe distance. We recall that in the Fe/S
liquid, the Fe-S distance was $\approx 1.95$ \AA~ (Alf\`e and Gillan,
1998a). An important feature of the liquid is that each oxygen has on
average only 9 iron neighbours, whereas it has between 4 and 5 oxygen
neighbours. This atomic environment of oxygen is very different from
that produced by the cubic structure of Fe$_3$O. By contrast in the
BiI$_3$ structure of Fe$_3$O oxygen has 11 iron neighbours and 2
oxygen neighbours. The dilemma in making crystal structures for Fe/O
solid solutions at core pressures is that we want to achieve close
packing because of the high pressure, but the atoms that are being
packed have different sizes. It seems that the cubic structure of
Fe$_3$O is not a good solution.  The BiI$_3$ structure is better, even
though it means putting oxygen atoms next each other.  There may be
yet better ways.

Our analysis of the electronic structure of the liquid gives further
insight. Here, the important feature is that Fe-O bonding is only
partially ionic, with a substantial covalent contribution. The
implication that there is partial electron transfer from Fe to O is
relevant, because presumably if O carried a full ionic charge of
$-2|e|$ oxygen atoms would be more reluctant to become neighbours of
each other. The electronic structure shows that there is no
detectable covalent interaction between oxygens, so the fact that they
become neighbours cannot be attributed to covalency. 

Finally, we have studied the diffusion coefficient of iron and oxygen
atoms and the viscosity of the liquid at $33 \%$ composition. The
finding is that both species diffuse more rapidly than in either pure
liquid iron or the liquid Fe/S alloy: the diffusion coefficients have
roughly twice the value that they have in pure Fe and Fe/S at the same
pressure and temperature. We also find that the viscosity of the Fe/O
liquid is about half what it is in those other systems. This means
that if the major light element in the outer core was oxygen, our
earlier conclusion (de Wijs et al., 1998) about the low viscosity of
the outer core would be confirmed, and indeed strengthened.

\section{Conclusions}

We are led to the following conclusions: if the light impurity in the
outer core is mainly oxygen, then its molar concentration would have
to be $\approx 28 \%$; in this region of concentration, we have
strong evidence that the liquid is stable against phase separation;
the proposed miscibility is not in conflict with the large formation
enthalpies predicted by earlier {\em ab initio} calculations, because
those calculations were based on assumed structures that are not the
most stable; the proposed Fe/O liquid alloy has an even lower
viscosity than that of pure Fe and the relevant Fe/S alloy under the
same conditions.

\section*{Acknowledgments} 

The work of DA is supported by NERC grant GST/O2/1454 to G. D. Price
and M. J. Gillan. We thank the High Performance Computing Initiative
for allocations of time on the Cray T3D and T3E at Edinburgh Parallel
Computer Centre, these allocations being provided through the Minerals
Physics Consortium (GST/02/1002) and the U.K.  Car-Parrinello
Consortium. We thank Dr. G. Kresse and Dr. G. de Wijs for valuable
technical assistance, and Dr. L. Vo\v{c}adlo
and Dr. D. Sherman for useful discussions. We thank Dr. W. Temmerman
for useful advice about the LMTO-46 code, and the Psi-k Network for
making the code available.

\newpage
\section*{References}

{\small
\medskip\noindent Alder, B. J., 1966. Is the mantle soluble in the
core?  J. Geophys. Res., 71: 4973-4979.

\medskip\noindent Alf\`e, D., and Gillan, M. J., 1998a. First principles
simulations of liquid Fe-S under Earth's core
conditions. Phys. Rev. B, {\it in press}.

\medskip\noindent Alf\`e, D., and Gillan, M. J., 1998b. The first
principles calculation of transport coefficient. Phys. Rev. Lett.,
{\it submitted}.

\medskip\noindent Allen, M. P., Tildesley, D. J., 1987. Computer
Simulation of Liquids. Oxford University Press, Oxford.

\medskip\noindent Bhatia, A. B., Thornton, D. E., 1970. Structural
aspects of the electrical resistivity of binary alloys. Phys. Rev. B,
2: 3004-3012.

\medskip\noindent Birch, F., 1952. Elasticity and composition of the
Earth's interior. J. Geophys. Res., 57: 227-286.

\medskip\noindent Birch, F., 1964. Density and composition of mantle
and core.  J. Geophys. Res., 69: 4377-4388.

\medskip\noindent Boehler, R., 1993. Temperature in the Earth's core
from melting-point measurements of iron at high static
pressures. Nature. 363: 534-536.

\medskip\noindent Bullen, K. E., 1973. Cores of terrestrial planets. Nature,
243: 68-70.

\medskip\noindent Car, R., and Parrinello, M., 1985.  Unified approach
for molecular-dynamics and density-functional theory.
Phys. Rev. Lett., 55: 2471-2474.

\medskip\noindent Clark, S. P., 1963. Variation of density in the
Earth and the melting curve in the mantle.  In: The Earth Sciences,
edited by T. W. Donnelly, University of Chicago Press, Chicago.

\medskip\noindent Distin, P. A., Whiteway S., and Masson, C., 1971.
Solubility of oxygen in liquid iron from 1785$^{\rm o}$ to 1960$^{\rm
o}$.  A new technique for the study of slag-metal
equilibria. Canad. Metallurg. Quart., 10: 13-18.

\medskip\noindent Dubrovskiy, V. A. and Pan'kov, V. L., 1972. On the
composition of the Earth's core. Acad. Sci. USSR Phys. Solid Earth, 7:
452-455.

\medskip\noindent Fischer, W. A., and Schumacher, J. F., 1978. Die
S\"attigungsl\"oslichkeit von Reineisen an Saurstoff vom Schmelzpunkt
bis 2046$^{\rm o}$ C, ermittelt mit dem Schwebeschmelzverfahren.
Arch. Eisenh\"uttenwes., 49: 431-435.

\medskip\noindent Friedel, J., 1969. Fundamental aspects in the bonding of
transition metals. In: The Physics of Metals, ed. J. Ziman, Cambridge
University Press, Cambridge, 494 pp.

\medskip\noindent Fukai, Y., and Akimoto, S., 1983. Hydrogen in the
Earth's core.  Proc. Jpn. Acad., Ser. B, 59: 158-162.

\medskip\noindent Gillan, M. J., 1997. The virtual matter
laboratory. Contemp. Phys., 38: 115-130.

\medskip\noindent Goarant, F., Guyot, F., Peyronneau, J., and Poirier,
J.-P., 1992. High-pressure and high-temperature reactions between
silicates and liquid iron alloys in the diamond anvil cell, studied
by analytical electron microscopy. J. Geophys. Res., 97: 4477-4487.

\medskip\noindent Hohenberg, P., and Kohn, W., 1964. Inhomogeneous
electron gas.  Phys. Rev. 136: B864-B871.

\medskip\noindent Holender, J. M., and Gillan, M. J., 1996.
Composition dependence of the structure and electronic properties of
liquid Ga-Se alloys studied by ab initio molecular dynamics simulation.
Phys. Rev. B, 53: 4399. 

\medskip\noindent Jones, R. O., and Gunnarsson, O., 1989. The density
functional formalism, its applications and prospects. Rev. Mod. Phys.,
61: 689-746.

\medskip\noindent Kirchhoff, F., Holender, J. M., and Gillan, M. J.,
1996a.  The structure, dynamics and electronic structure of liquid
Ag-Se alloys investigated by ab initio simulation. Phys. Rev. B, 54: 190-202.

\medskip\noindent Kirchhoff, F., Gillan, M. J., Holender, J. M.,
Kresse, G., and Hafner, J., 1996b. Structure and bonding in liquid Se.
J. Phys. Condens. Matter, 8: 9353-9357.

\medskip\noindent Knittle, E., Jealnoz, R., 1991. Earth's core-mantle
boundary: results of experiments at high pressures and
temperatures. Science, 251: 1438-1443.

\medskip\noindent  Kohn, W., and Sham, L., 1965. Self-consistent equations 
including exchange and correlation effects. Phys. Rev. 140: A1133-A1138.

\medskip\noindent Kresse, G., and Hafner, J., 1993.
Ab-initio molecular-dynamics for open-shell transition-metals. Phys. Rev. B,
48: 13115-13118.

\medskip\noindent Kresse, G., and  Furthm\"{u}ller, J., 1996a.
Efficiency of ab-initio total-energy calculations for metals and 
semiconductors using a plane-wave basis-set.
Comput. Mater. Sci.,  6: 15-50.

\medskip\noindent Kresse, G., and Furthm\"{u}ller, J., 1996b.
Efficient iterative schemes for ab-initio total-energy calculations
using a plane-wave basis-set.  Phys. Rev. B, 54: 11169-11186.

\medskip\noindent Krier, G., Jepsen, O., Burkhardt, A., and Andersen,
O. K., 1994.  LMTO-46 code, Max-Planck-Institut f\"ur
Festk\"orperforschung, Stuttgart.

\medskip\noindent Lewis, J. S., 1973. Chemistry of the planets.
Annu. Rev. Phys. Chem., 24: 339-351.

\medskip\noindent Louie, S. G., Froyen, S., and Cohen, M. L., 1982.
Non-linear ionic pseudopotentials in spin-density-functional calculations.
Phys. Rev. B, 26: 1738-1742.

\medskip\noindent MacDonald, G. J. F., and Knopoff, L., 1958.
The chemical composition of the outer core. J. Geophys., 1: 1751-1756.

\medskip\noindent Mao, H. K., Wu, Y., Chen, L. C., Shu, J. F., and
Jephcoat, A. P., 1990.  Static compression of iron to 300 GPa and
Fe0.8Ni0.2 alloy to 260 GPa:  implications for composition of the
core.  J. Geophys. Res., 95: 21737-21742.

\medskip\noindent Mason, B., 1966. Composition of the Earth. Nature,
211: 616-618.

\medskip\noindent Monkhorst, H. J., and Pack, J. D., 1976. Special
points for Brillouin-zone integrations. Phys. Rev. B, 13: 5188-5192.

\medskip\noindent Murthy, V., and Hall, H. T., 1970. On the possible
presence of sulfur in the Earth's core. Phys. Earth Planet. Inter., 2:
276-282.

\medskip\noindent  Nos\'e, S., 1984. A molecular dynamics method for 
simulations in the canonical ensemble. 
Molec. Phys.,  52: 255-268.

\medskip\noindent Ohtani, E., and Ringwood, A. E., 1984 Composition
of the core, I.  Solubility of oxygen in molten iron at high
temperatures.  Earth Planet. Sci. Lett., 71: 85-93.

\medskip\noindent Ohtani, E., Ringwood, A. E. and Hibberson, W.,
1984. Composition of the core, II. Effect of high pressure on
solubility of FeO in molten iron.  Earth Planet. Sci. Lett., 71:
94-103.

\medskip\noindent Parr, R. G., and Yang, W., 1989. Density-Functional
Theory of Atoms and Molecules, Oxford University Press, Oxford.

\medskip\noindent Poirier, J.-P., 1991.  Introduction to the Physics
of the Earth's Interior. Cambridge University Press, Cambridge.

\medskip\noindent Poirier, J.-P., 1994. Light elements in the Earth's
outer core: A critical review. Phys. Earth Planet. Interiors. 85:
319-337.

\medskip\noindent Pulay, P., 1980. Convergence acceleration of
iterative sequences. The case of SCF iteration. Chem. Phys. Lett., 73:
393.

\medskip\noindent Ringwood, A. E., 1959. On the chemical evolution and
densities of the planets. Geochem. J., 15: 257-283.

\medskip\noindent Ringwood, A. E., 1961. Silicon in the metal phase of
enstatite chondrites and some geochemical
implications. Geochim. Cosmochim. Acta, 25: 1-13.

\medskip\noindent Ringwood, A. E., 1966. Chemical evolution of the
terrestrial planets. Geochim. Cosmochim. Acta, 30: 41-104.

\medskip\noindent Ringwood, A. E., 1977. On the composition of the
core and implications for the origin of the Earth.
Geochim. Cosmochim. Acta, 11: 111-135.

\medskip\noindent Ringwood, A. E., and Hibberson, W., 1990. The system
Fe-FeO revisited. Phys. Chem. Minerals, 17: 313-319.

\medskip\noindent Secco, R. A. 1995. Viscosity of the Outer Core. 
In: Mineral Physics and
Crystallography: A Handbook of Physical Constants, ed. T. J. Ahrens,
Americal Geophysical Union, 218 pp.

\medskip\noindent Sherman, D. M., 1995. Stability of possible Fe-FeS
and Fe-FeO alloy phases at high pressure and the composition of the
Earth's core.  Earth Planet. Sci. Lett., 132: 87-98.

\medskip\noindent  Shimoji, M., and Itami, T., 1986. Atomic Transport in
Liquid Metals, Trans Tech Publications, Aedermannsdorf, p. 191.

\medskip\noindent S\"oderlind, P., Moriarty, J. A., and Wills, J. M.,
1996.  First principles theory of iron up to the Earth's core
pressures: Structural, vibrational, and elastic properties.
Phys. Rev. B, 53: 14063-14072.

\medskip\noindent \v{S}tich, I., Car, R., and Parrinello, M., 1989.
Bonding and disorder in liquid silicon. Phys. Rev. Lett., 63:
2240-2243.

\medskip\noindent Stixrude, L., Cohen, R. E., and Singh, D. J., 1994.
Iron at high pressure: Linearized-augmented-plane-wave computation in
the generalized-gradient approximation.  Phys. Rev. B, 50: 6442-6445.

\medskip\noindent Sugino, O., and Car, R., 1995.  Ab-initio
molecular-dynamics study of first-order phase-transitions.  Melting of
silicon.  Phys. Rev. Lett., 74: 1823-1826.

\medskip\noindent Suzuki, T., Akimoto S., and Yagi, T.,
1989. Metal-silicate-water reaction under high pressure, I, Formation
of metal hydride and implications for composition of the core and
mantle. Phys. Earth Planet. Inter., 56: 377-388.

\medskip\noindent Urey, H. C., 1960. On the chemical evolution and
densities of the planets. Geochim. Cosmochim. Acta, 18: 151-153.

\medskip\noindent Vanderbilt, D., 1990. Soft self-consistent
pseudopotentials in a generalized eigenvalue formalism. Phys. Rev. B,
41: 7892-7895.

\medskip\noindent Verlet, L., 1967. Computer `experiments' on
classical fluids.  I. Thermodynamical properties of Lennard-Jones
molecules.  Phys. Rev., 159: 98-103.

\medskip\noindent Vo\v{c}adlo, L., de Wijs, G. A., Kresse, G., Gillan,
M. J., and Price, G. D., 1997. First principles calculations on
crystalline and liquid iron at earth's core conditions. Faraday
Discuss., 106: 205-217.

\medskip\noindent Wang, Y., and Perdew, J., 1991. Correlation hole of
the spin-polarized electron gas, with exact small-wave-vector and
high-density scaling. Phys. Rev. B, 44: 13298-13307.

\medskip\noindent de Wijs, G. A., Kresse, G., and Gillan, M. J., 1998.
First order phase transitions by first-principles free energy
calculations: The melting of Al.  Phys. Rev. B, 57: 8223-8234.

\medskip\noindent de Wijs, G. A., Kresse, G., Vo\v{c}adlo, L., Dobson,
D., Alf\`e, D., Gillan, M. J., Price, G. D., 1998. The viscosity of
liquid iron at the physical conditions of the Earth's core. Nature,
392: 805-807.

\medskip\noindent Wood, B. J. 1993. Carbon in the core. Earth
Planet. Sci. Lett., 117: 593-607.

\medskip\noindent Wyckoff, R. W. G., 1964. Crystal Structures, 2nd
edition, Vol. 2, Interscience, New York.  }

\newpage

\begin{tabular}{lcccc}
\hline \hline
& & PP(spin-unrestricted) & PP(spin-restricted) & FLAPW \\
\hline
Fe($\epsilon$)  & $\rho_0$    &  8910   &  8910 & 8780 \\
         & $K$               &  283   &  283 & 260  \\
         & $K'$                &  4.39  &  4.39 & 4.53 \\
& & & & \\
FeO(B8)  & $\rho_0$    &  5650  &  6980 & 5810 \\
         & $K$               &  92   &  258 & 173    \\
         & $K'$                  &  4.96  &  4.4 & 2.93   \\
         &$\mu$                 &  2.0  &  & not reported        \\
& & & & \\
Fe$_3$O(cubic)  & $\rho_0$    &  6500  &  7550 & 7420 \\
         & $K$               &  123  & 226 & 223   \\
         & $K'$                   &  4.23  &  4.04 & 4.02   \\
         &$\mu$                 &  2.34  &   & small       \\
& & & & \\
Fe$_4$O  & $\rho_0$    & 6880   &  7840 & 7830 \\
         & $K$              & 135   & 273 & 310  \\
         & $K'$                  & 4.82   &  4.26 & 4.17   \\
         &$\mu$              &  2.0  &  &            \\
& & & & \\
Fe$_3$O(BiI$_3$)  & $\rho_0$ &    &    7930 & \\
         & $K$            &    &    248  & \\
         & $K'$                &    &    4.29  & \\
\hline \hline
\end{tabular}

\bigskip
\noindent Table 1: Calculated equilibrium properties of crystals in the Fe/O
system: equilibrium mass density $\rho_0$ (kg~m$^{-3}$), bulk modulus
$K$ (GPa), the pressure derivative $K'=dK/dP$, and magnetic moment per
atom $\mu$ (units of Bohr magneton). Results are given for the present spin
unrestricted and restricted pseudopotential (PP) calculations and the
FLAPW calculations of Sherman (1995). The structures of
the cubic Fe$_3$O, Fe$_4$O and BiI$_3$-structure Fe$_3$O are shown in
Fig. 1.

\bigskip
\bigskip
\begin{tabular}{ccccccc}
\hline \hline
Volume (\AA$^3$) & Pressure (GPa) & $b/a$ & $c/a$ & $\alpha$ &
$\beta$ & $\gamma$ \\
\hline 
22 & 467 & 1.01016 & 0.95443  & 123.76 &  90.70 &  79.04 \\
24 & 325 & 1.00988 & 0.95322  & 123.82 &  90.88 &  78.79 \\
26 & 226 & 1.01010 & 0.95164  & 123.85 &  91.03 &  78.55 \\
28 & 156 & 1.00772 & 0.94984  & 123.89 &  91.26 &  78.28 \\
30 & 105 & 1.00574 & 0.94874  & 123.85 &  91.39 &  78.07 \\
32 & 67  & 0.99711 & 0.94465  & 123.82 &  91.79 &  77.69 \\
34 & 40  & 0.99452 & 0.94303  & 123.76 &  91.97 &  77.37 \\
36 & 19  & 0.99193 & 0.94258  & 123.57 &  91.98 &  76.95 \\
\hline \hline
\end{tabular}

\bigskip
\noindent Table 2: Cell parameters and pressure $P$ of Fe$_3$O in the BiI$_3$
structure (see Fig. 1) calculated at a series of volumes
(per Fe$_3$O unit).  The quantities $a$, $b$, $c$ are the magnitudes of
the primitive translation vectors, and $\alpha, \beta, \gamma$ are the
angles between the pairs $(a,b), (a,c)$ and $(b,c)$ respectively.

\newpage
\section*{Figure captions}

\medskip\noindent{\bf Fig. 1:} The crystal structures of cubic Fe$_3$O (left),
cubic Fe$_4$O (centre) and the BiI$_3$ form of Fe$_3$O (right) used to
calculate the formation enthalpies of Fe/O solid solutions. Large and
small spheres represent iron and oxygen respectively.

\medskip\noindent{\bf Fig. 2:} Calculated enthalpies of formation (per
formula unit) of solid solutions having compositions Fe$_3$O and
Fe$_4$O. Left panel shows spin-restricted results from present work
compared with FLAPW results of Sherman (1995); right panel shows
present spin-unrestricted results. Key to style of curves: present
cubic Fe$_3$O \protect\rule[.5mm]{5mm}{1mm}, cubic Fe$_3$O of Sherman
(1995) \protect\rule[1mm]{5mm}{0.3mm}, present cubic Fe$_4$O
\protect\rule[.5mm]{1.5mm}{1mm} \protect\rule[.5mm]{1.5mm}{1mm}
\protect\rule[.5mm]{1.5mm}{1mm}, cubic Fe$_4$O of Sherman (1995)
\protect\rule[1mm]{1.5mm}{0.3mm} \protect\rule[1mm]{1.5mm}{0.3mm}
\protect\rule[1mm]{1.5mm}{0.3mm}, present Fe$_3$O in the BiI$_3$
structure \protect\rule[1mm]{1.5mm}{.3mm}
\protect\rule[1mm]{.3mm}{.3mm} \protect\rule[1mm]{1.5mm}{.3mm}
\protect\rule[1mm]{.3mm}{.3mm} \protect\rule[1mm]{1.5mm}{.3mm}
\protect\rule[1mm]{.3mm}{.3mm}.  The isolated point shows formation
enthalpy of the amorphous structure obtained by quenching the liquid
(see Sec. \ref{liquid}).

\medskip\noindent{\bf Fig. 3:} Radial distribution functions
$g_{\alpha\beta}(r)$ obtained from simulation of liquid Fe/O at oxygen
molar concentration of $33 \%$.

\medskip\noindent{\bf Fig. 4:} The iron-iron radial distribution function
$g_{\rm FeFe}(r)$ from the present simulation of liquid Fe/O (oxygen
molar concentration of $33 \%$) compared with $g_{\rm FeFe}(r)$ from
simulation of pure liquid iron at similar pressure and temperature
(Vo\v{c}adlo et al., 1997).

\medskip\noindent{\bf Fig. 5:} Partial structure factors $S_{\alpha\beta}(k)$
calculated from simulation of liquid Fe/O at oxygen molar
concentration of $33 \%$.

\medskip\noindent{\bf Fig. 6:} Probability distribution $P_{\rm
FeO}(n,r^c_{\rm FeO})$ for number $n$ of oxygen neighbours of an iron
atom calculated from simulation of liquid Fe/O at oxygen molar
concentration of $33 \%$.

\medskip\noindent{\bf Fig. 7:} Snapshots of the simulated liquid Fe/O system
along three principal Cartesian axes. Top three panels show a
configuration from the period when phase separation was artificially
induced by application of an external force. Bottom three panels show
a configuration from the later period after removal of the external
force. Large and small spheres represent iron and oxygen
respectively.

\medskip\noindent{\bf Fig. 8:} Probability distribution $P_{\rm
FeO}(n,r^c_{\rm FeO})$ for number $n$ of oxygen neighbours of an iron
atom calculated from simulation of liquid Fe/O at oxygen molar
concentration of $25 \%$.  Results are average values for three
windows of length 0.1 ps at times of 0 ps, 0.5 ps and 1 ps after
removal of the external force used to induce phase
separation.

\medskip\noindent{\bf Fig. 9:} Time dependent diffusion coefficients
$D_\alpha(t)$ for iron and oxygen calculated from the simulation of
liquid Fe/O at oxygen molar concentration of $33 \%$.

\medskip\noindent{\bf Fig. 10:} Average over the five independent components of
the autocorrelation function of the traceless stress tensor $\phi(t)$
(left panel) and viscosity integral (solid curve) with its statistical
error (dashed curve) (right panel).

\medskip\noindent{\bf Fig. 11:} Electronic density of states (upper panel) and
local densities of states (lower panel) calculated for liquid Fe/O at
oxygen molar concentration of $25 \%$. Energy is referred to the Fermi
energy $E_f$.

\medskip\noindent{\bf Fig. 12:} Local densities of states for two selected
oxygen atoms taken from the simulation of liquid Fe/O at oxygen molar
concentration of $25 \%$. Atom O$_{\rm a}$ has 1 oxygen neighbour and
10 iron neighbours; atom O$_{\rm b}$ has 4 oxygen neighbours and 7
iron neighbours. Energy is referred to the Fermi energy $E_f$.

\newpage
\bigskip\centerline{FIGURE 1}
\bigskip\centerline{\psfig{figure=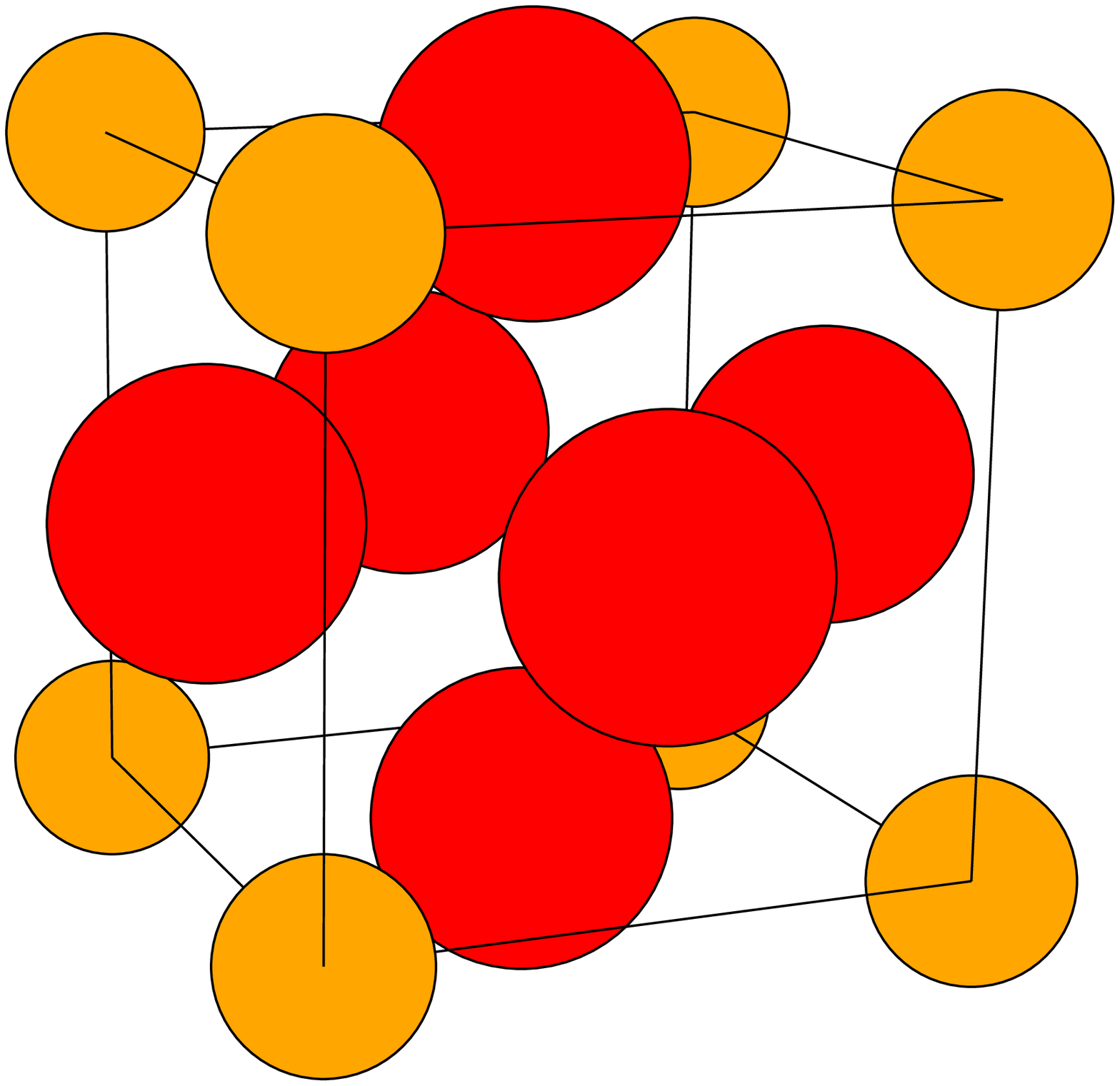,height=1.8in}\hskip 30pt
\psfig{figure=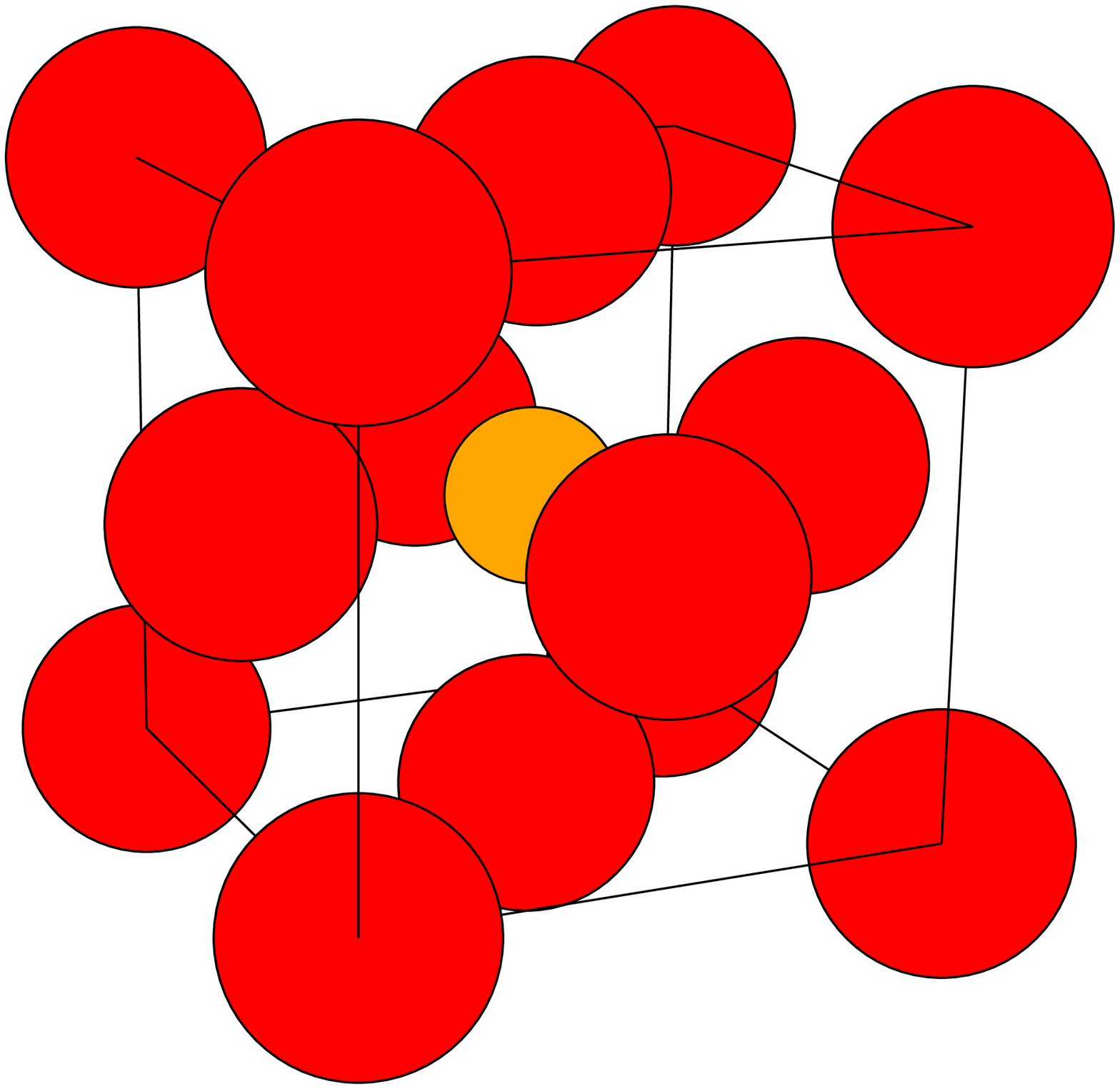,height=1.8in}\hskip 30 pt
\psfig{figure=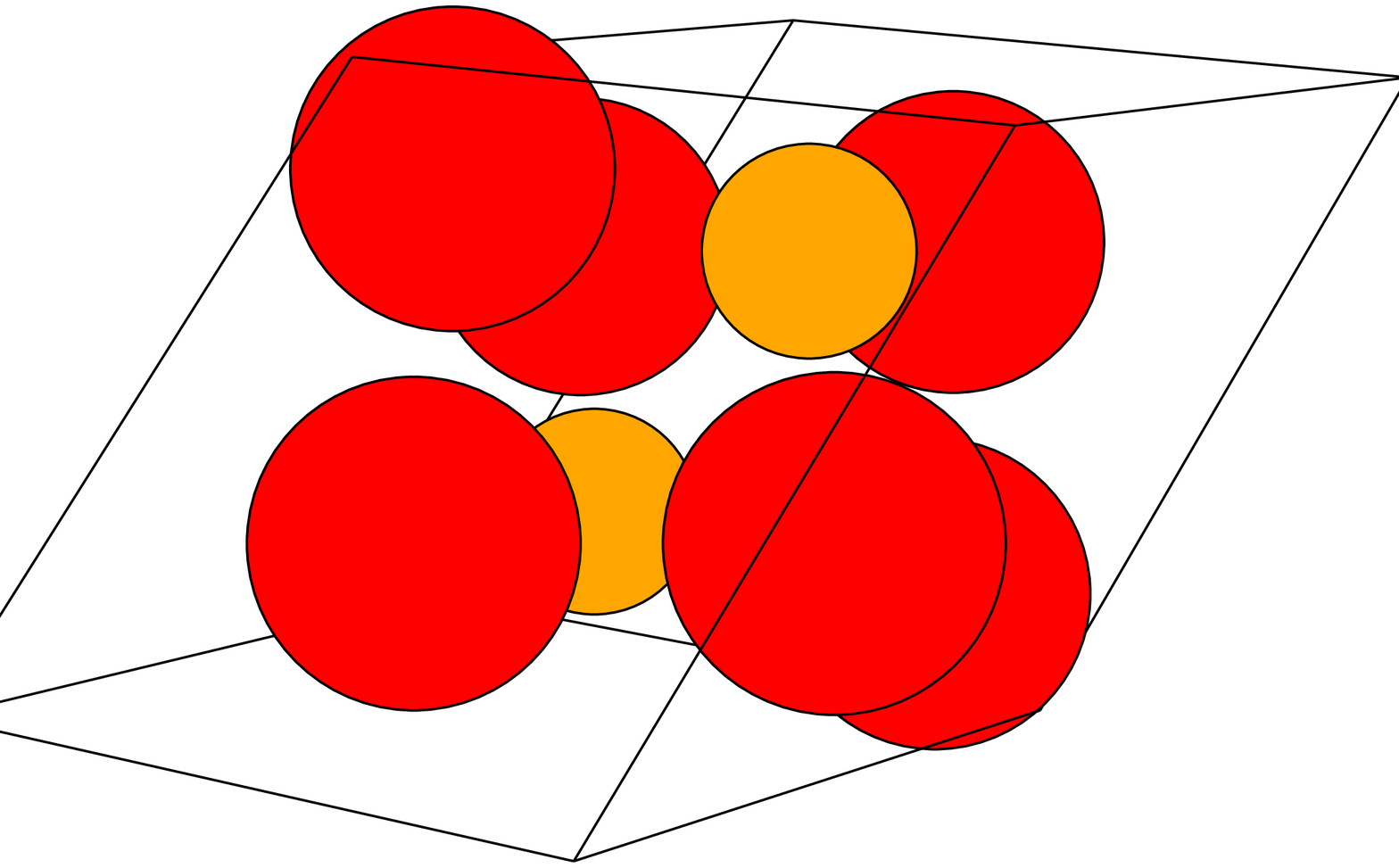,height=1.8in}}
 
\newpage
\bigskip\centerline{FIGURE 2}
\bigskip\centerline{\psfig{figure=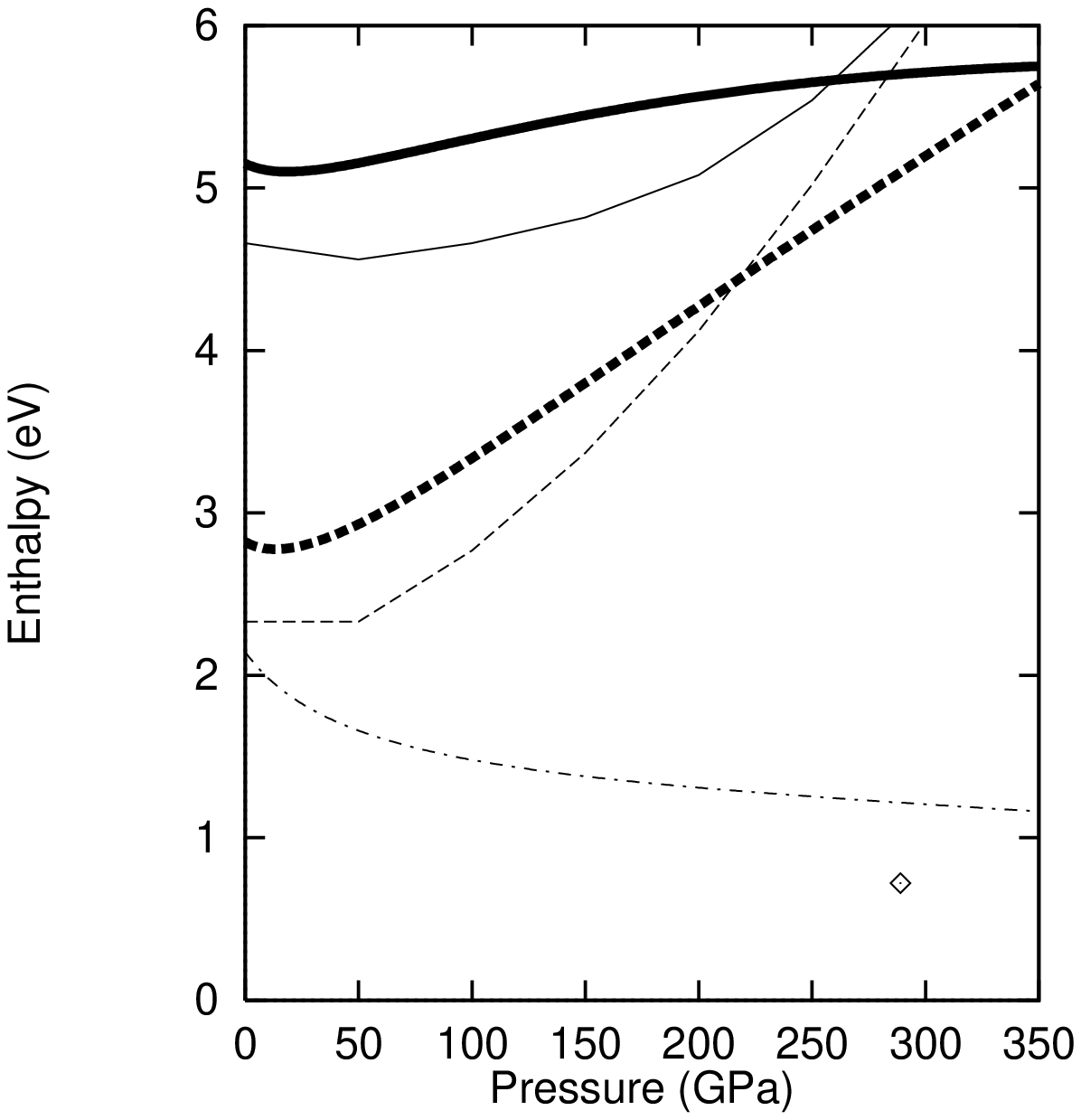,height=3in}\hskip 20pt
\psfig{figure=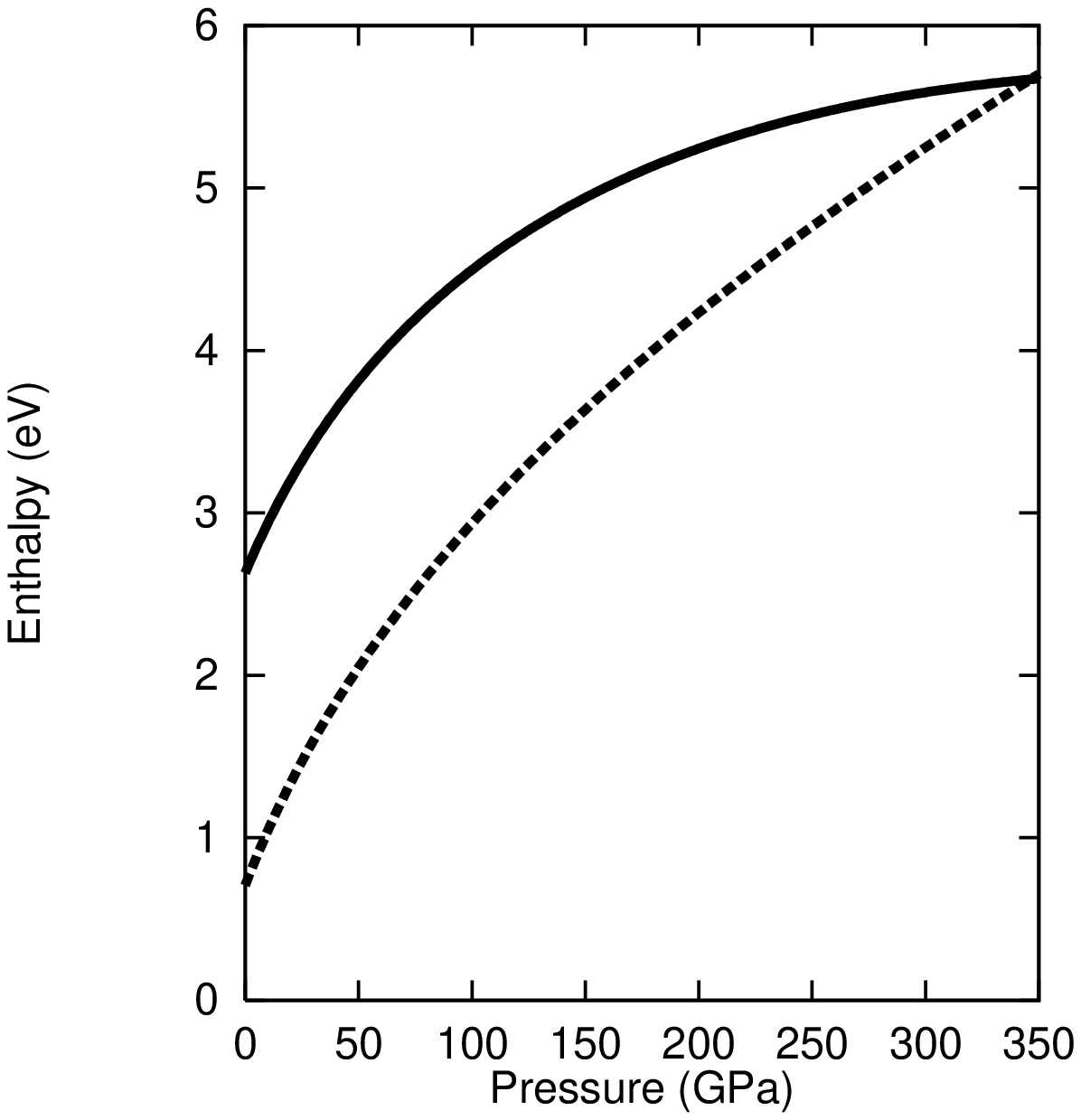,height=3in}}
 
\newpage
\bigskip\centerline{FIGURE 3}
\bigskip\centerline{\psfig{figure=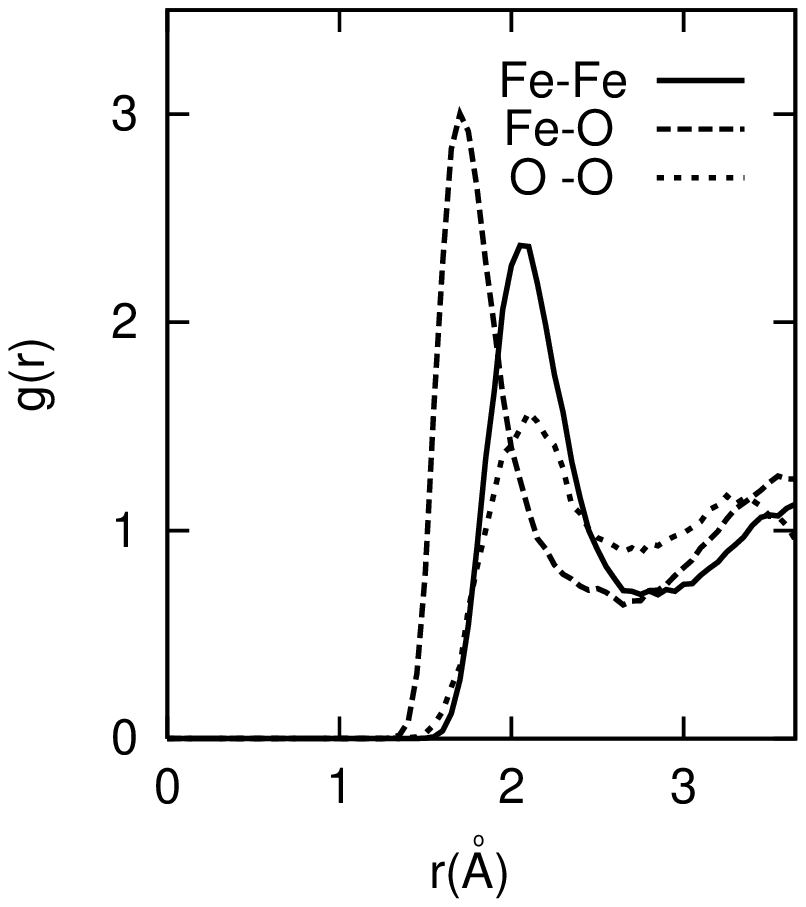,height=3.5in}}
 
\newpage
\bigskip\centerline{FIGURE 4}
\bigskip\centerline{\psfig{figure=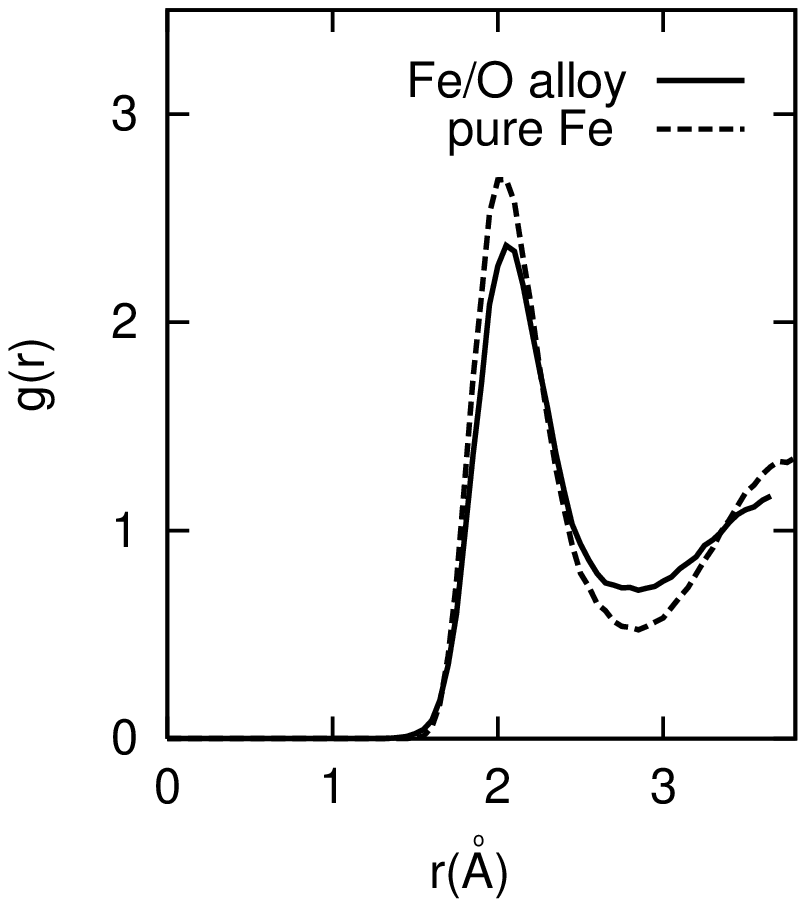,height=3.5in}}
 		
\newpage
\bigskip\centerline{FIGURE 5}
\bigskip\centerline{\psfig{figure=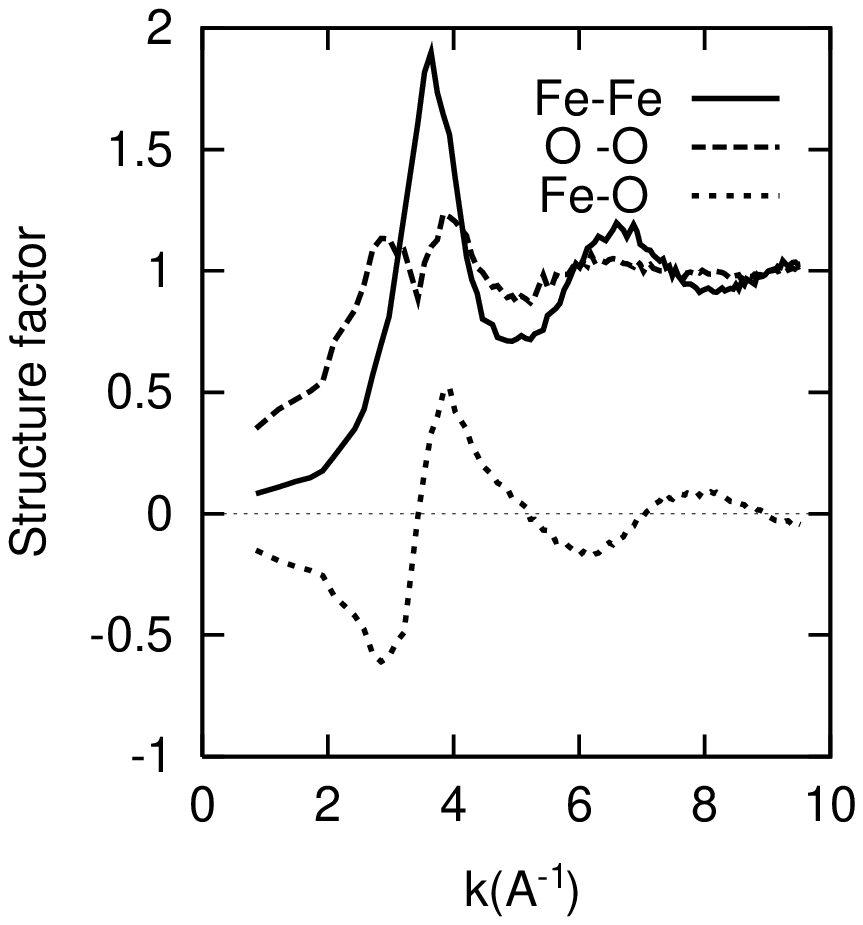,height=3.5in}}
 
\newpage
\bigskip\centerline{FIGURE 6}
\bigskip\centerline{\psfig{figure=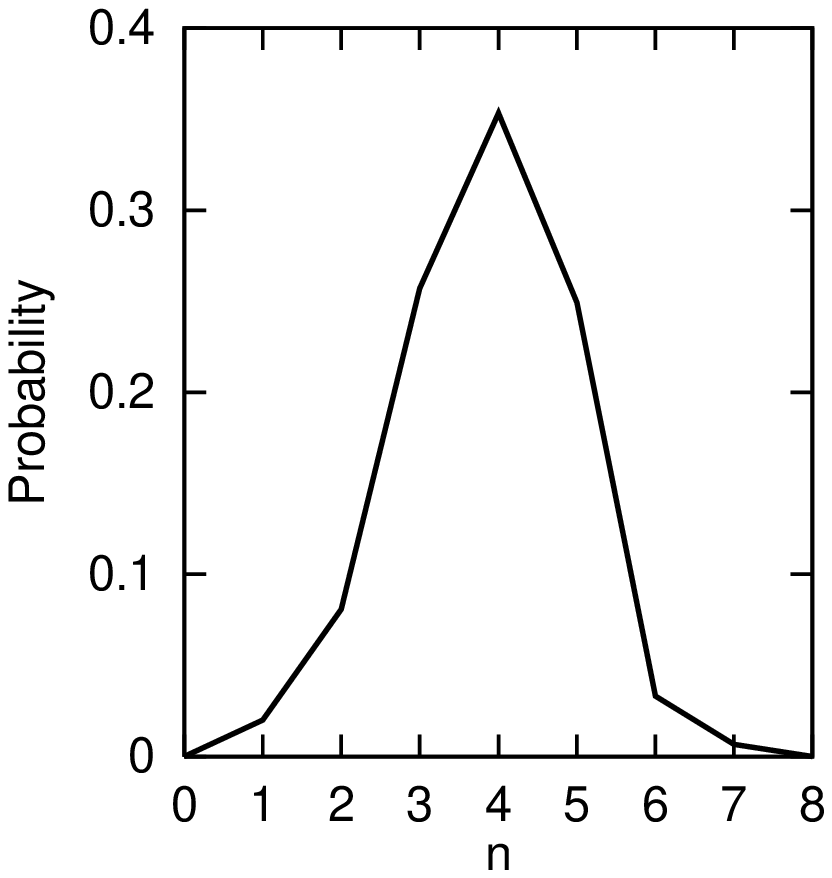,height=3.5in}}
 
\newpage
\bigskip\centerline{FIGURE 7}
\bigskip\centerline{\psfig{figure=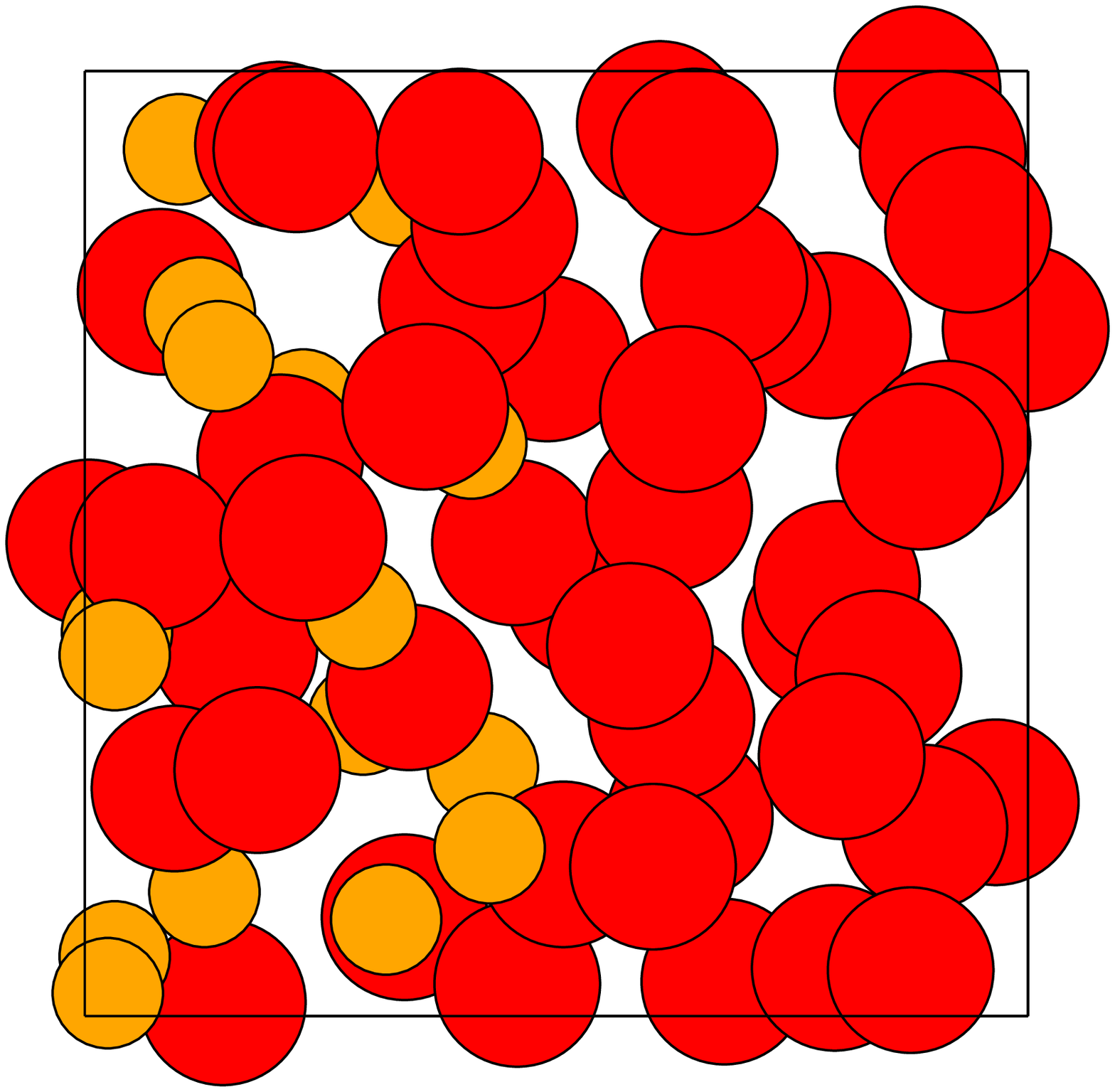,height=2.0in}\hskip 10pt
\psfig{figure=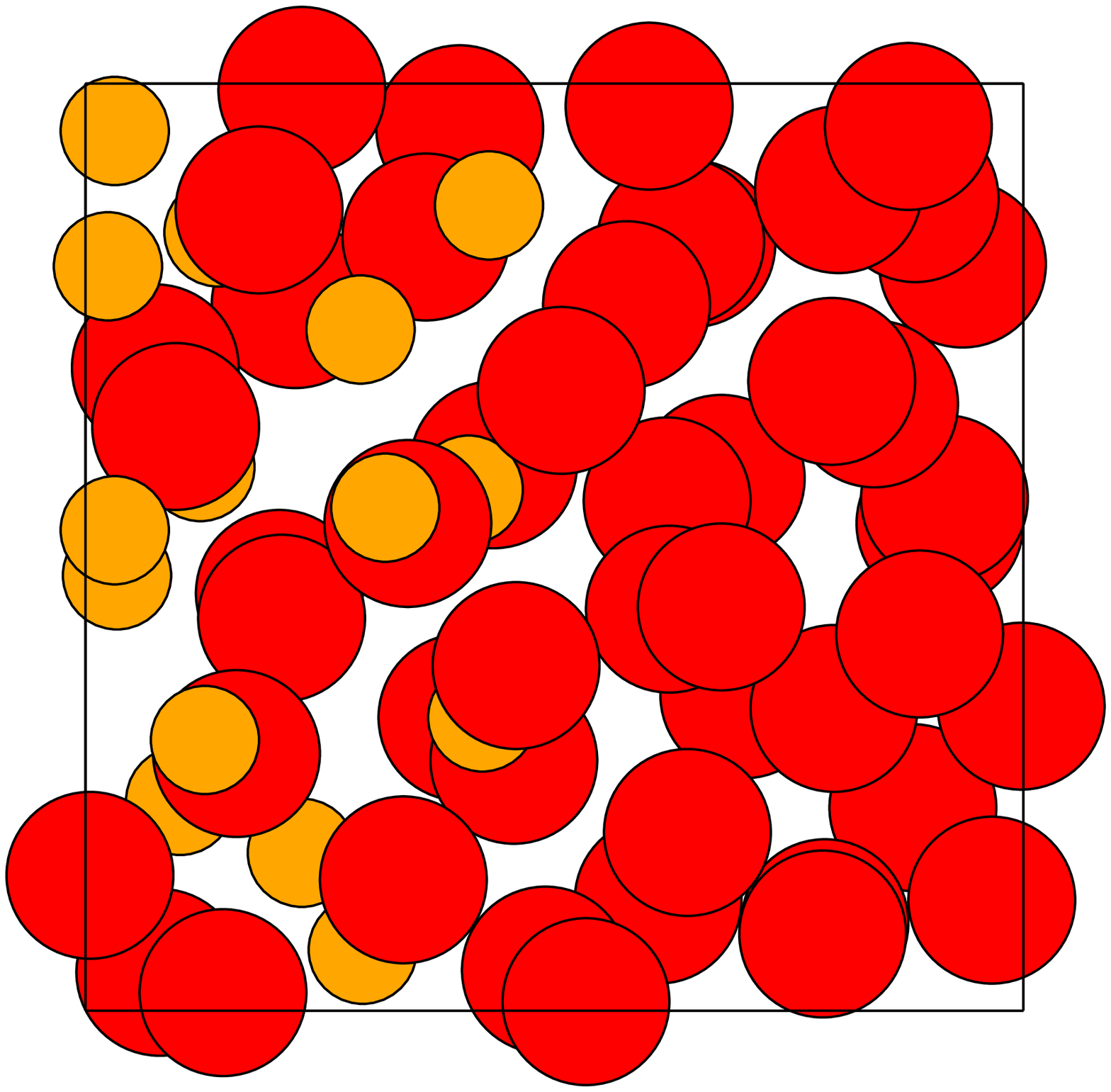,height=2.0in}\hskip 10pt 
\psfig{figure=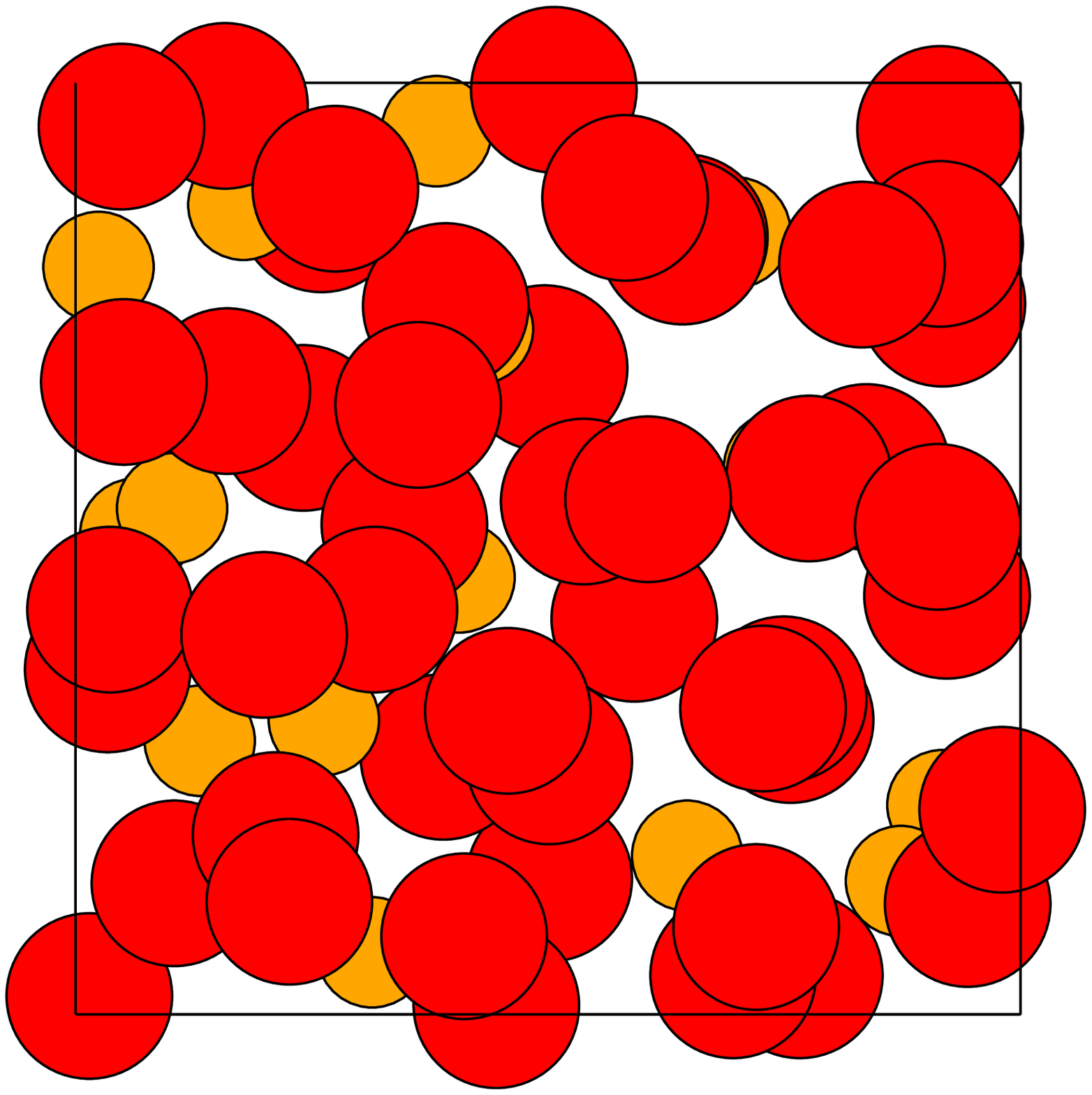,height=2.0in}}
\bigskip\centerline{\psfig{figure=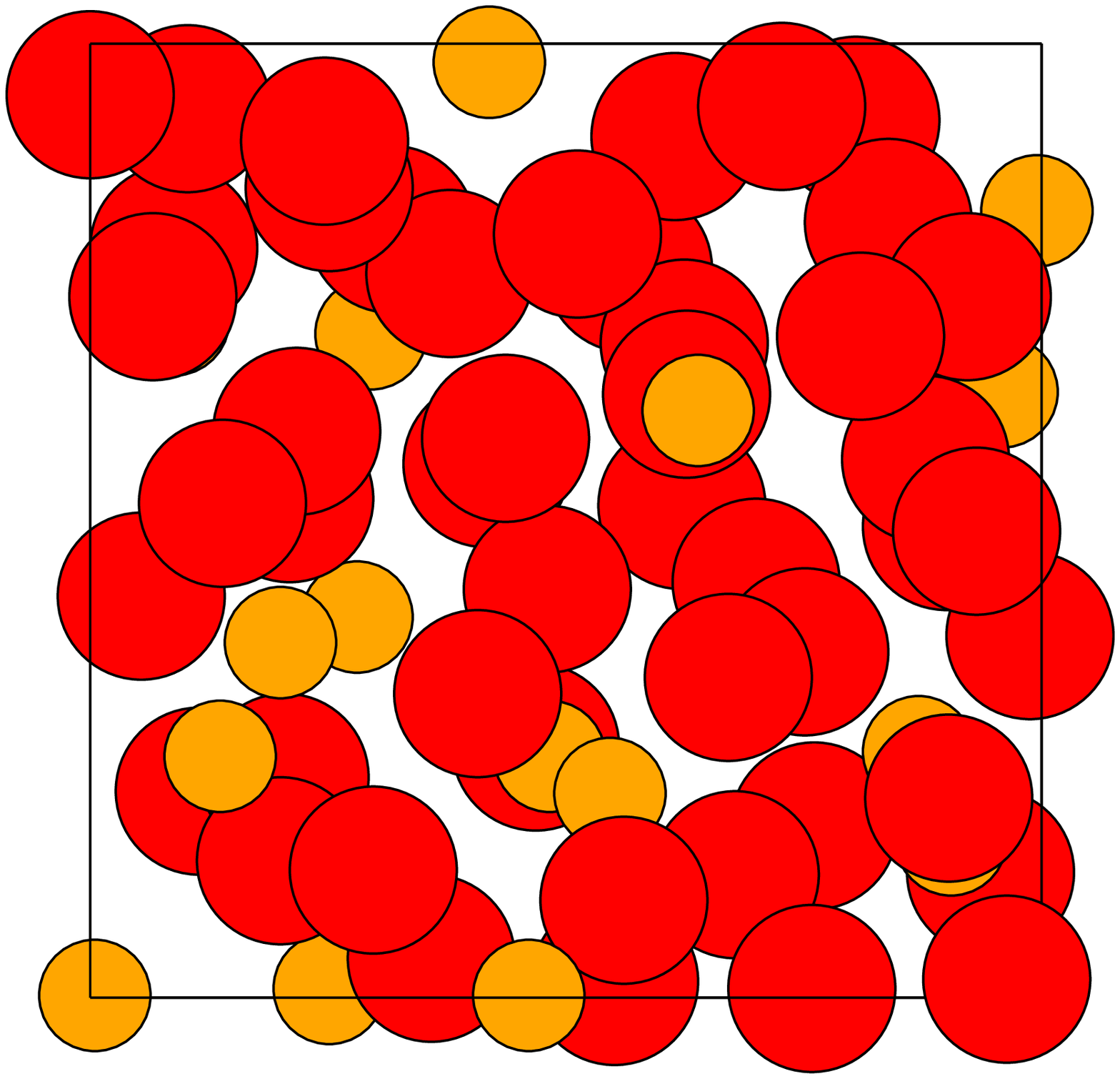,height=2.0in}\hskip 10pt
\psfig{figure=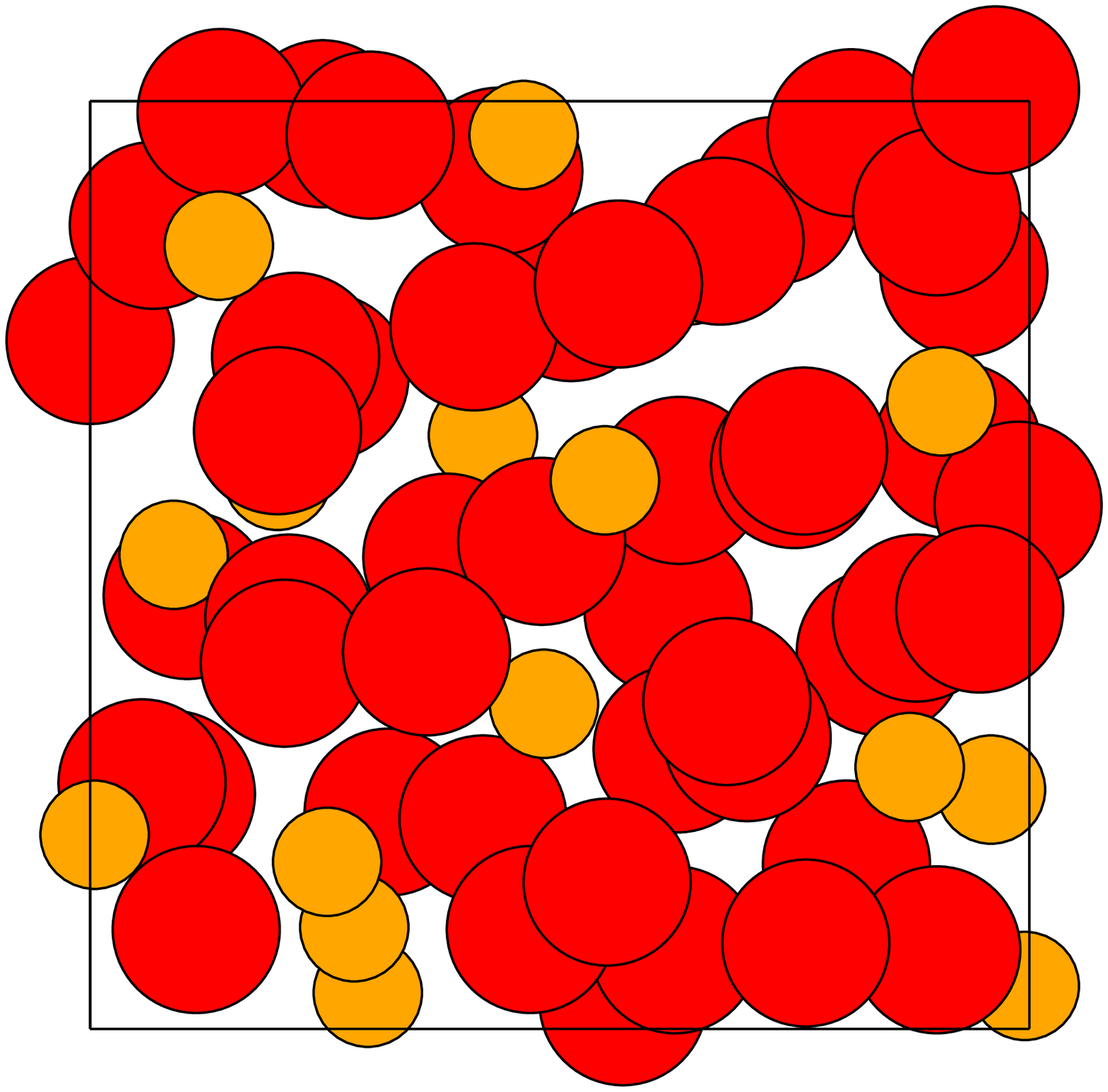,height=2.0in}\hskip 10pt 
\psfig{figure=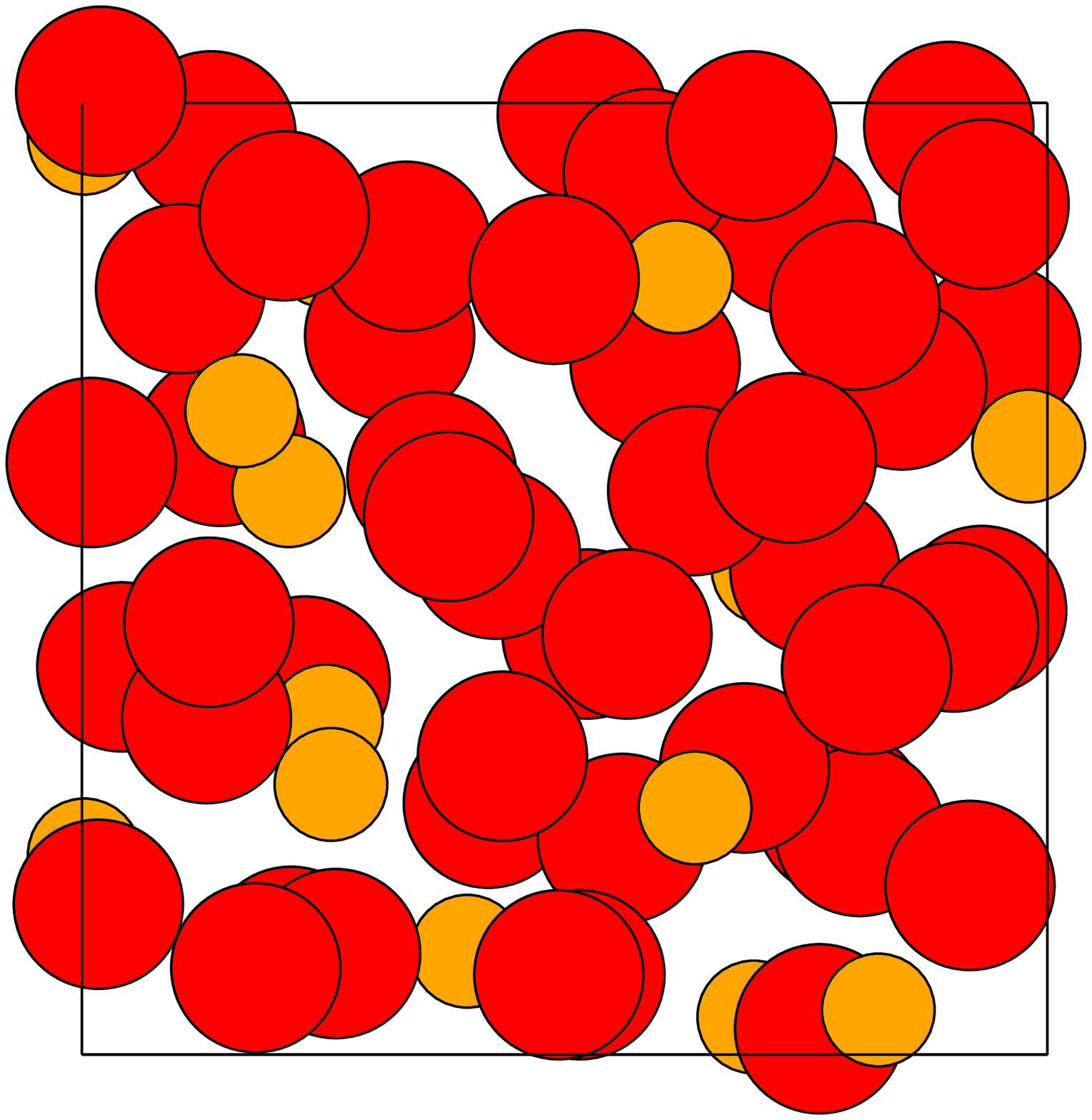,height=2.0in}}
 
\newpage
\bigskip\centerline{FIGURE 8}
\bigskip\centerline{\psfig{figure=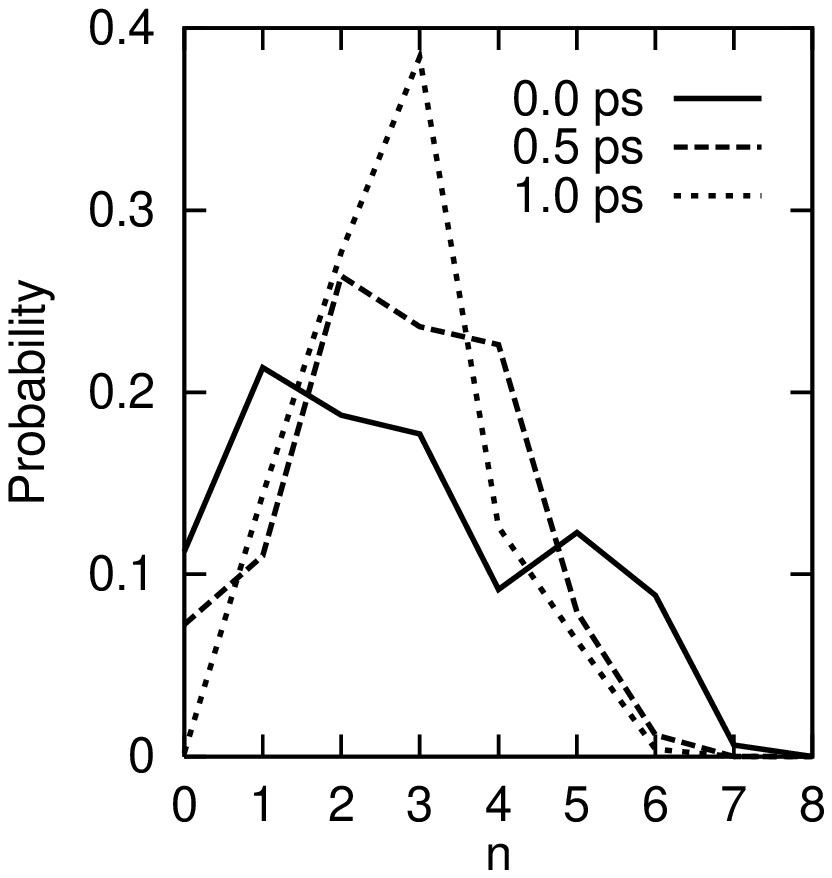,height=3.5in}}
 
\newpage
\bigskip\centerline{FIGURE 9}
\bigskip\centerline{\psfig{figure=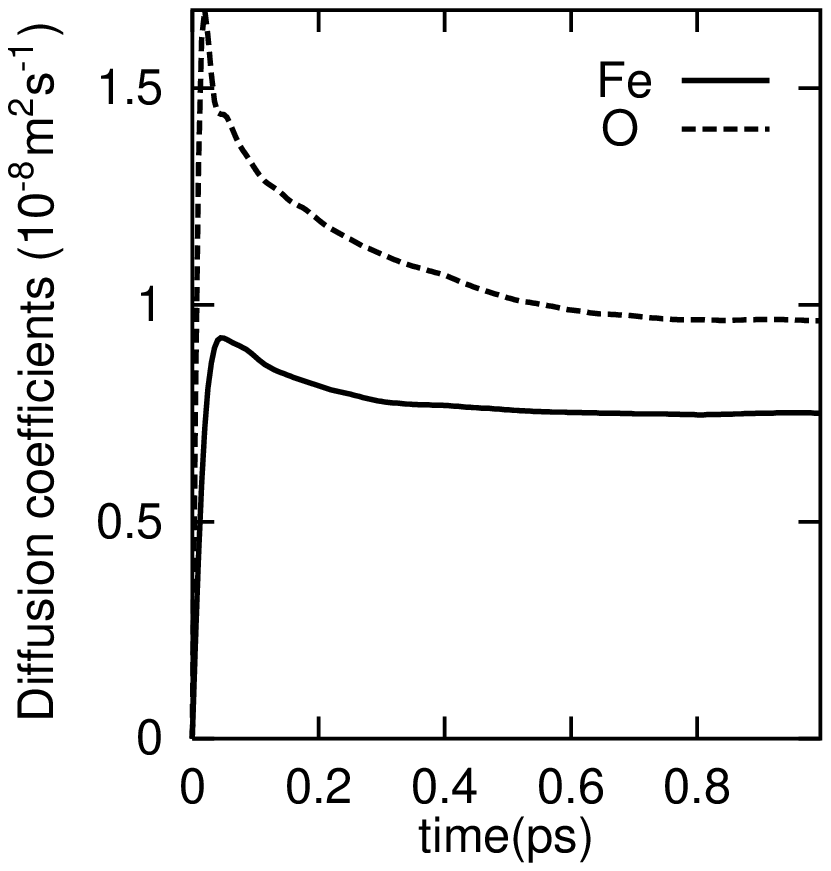,height=3.5in}}
 
\newpage
\bigskip\centerline{FIGURE 10}
\bigskip\centerline{\psfig{figure=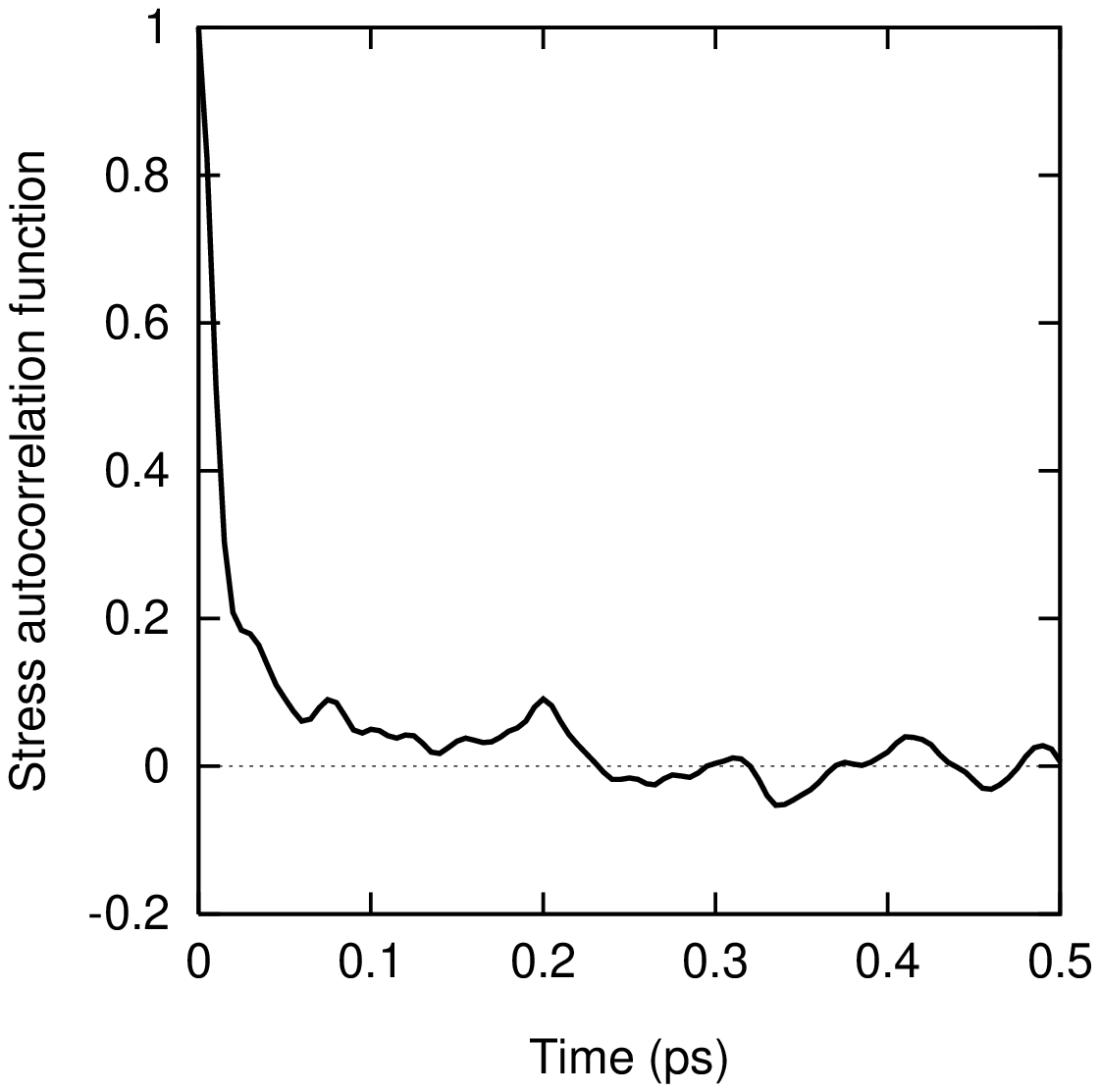,height=2.8in}\hskip 20pt
\psfig{figure=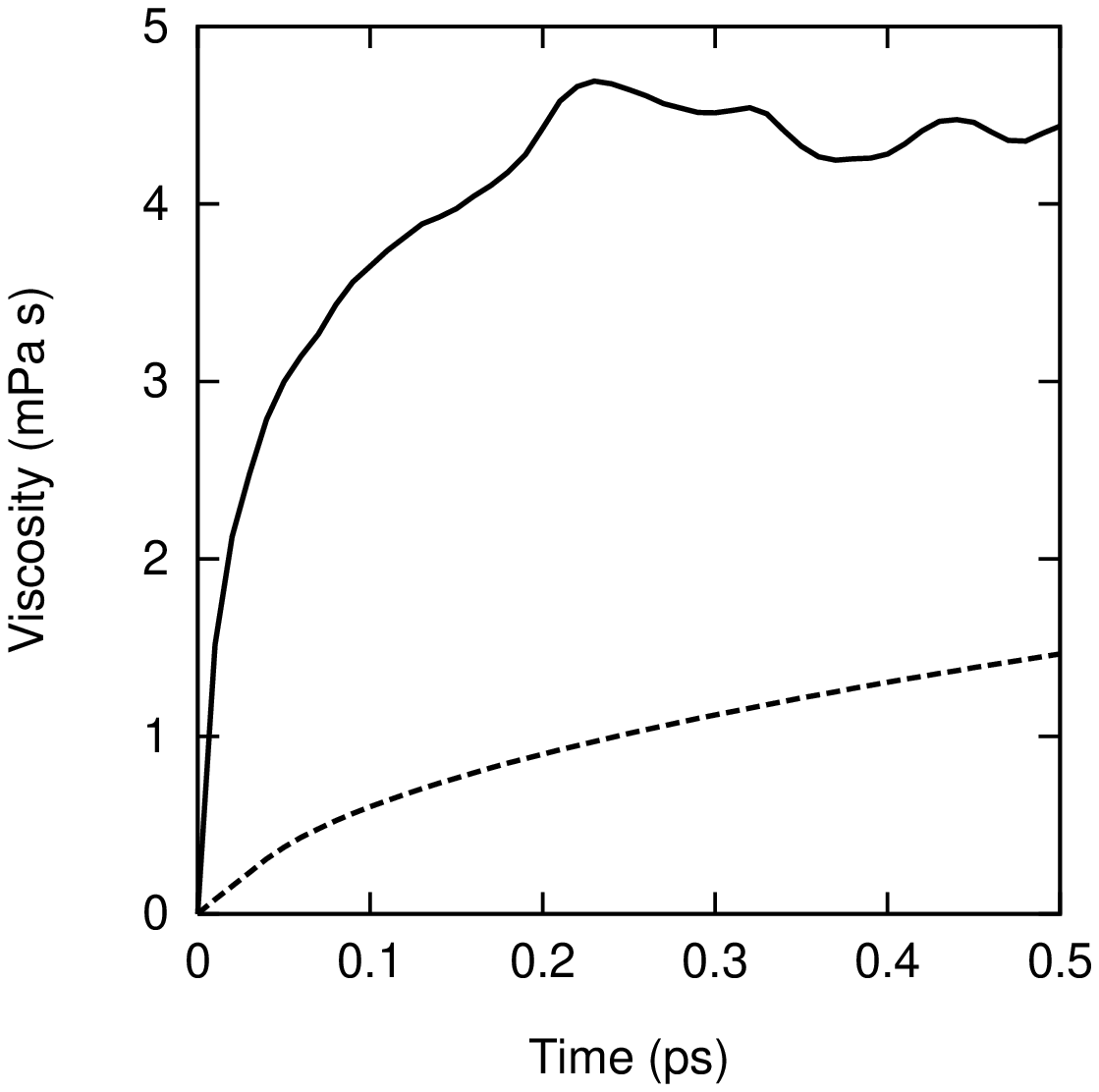,height=2.8in}} 
 
\newpage
\bigskip\centerline{FIGURE 11}
\bigskip\centerline{\psfig{figure=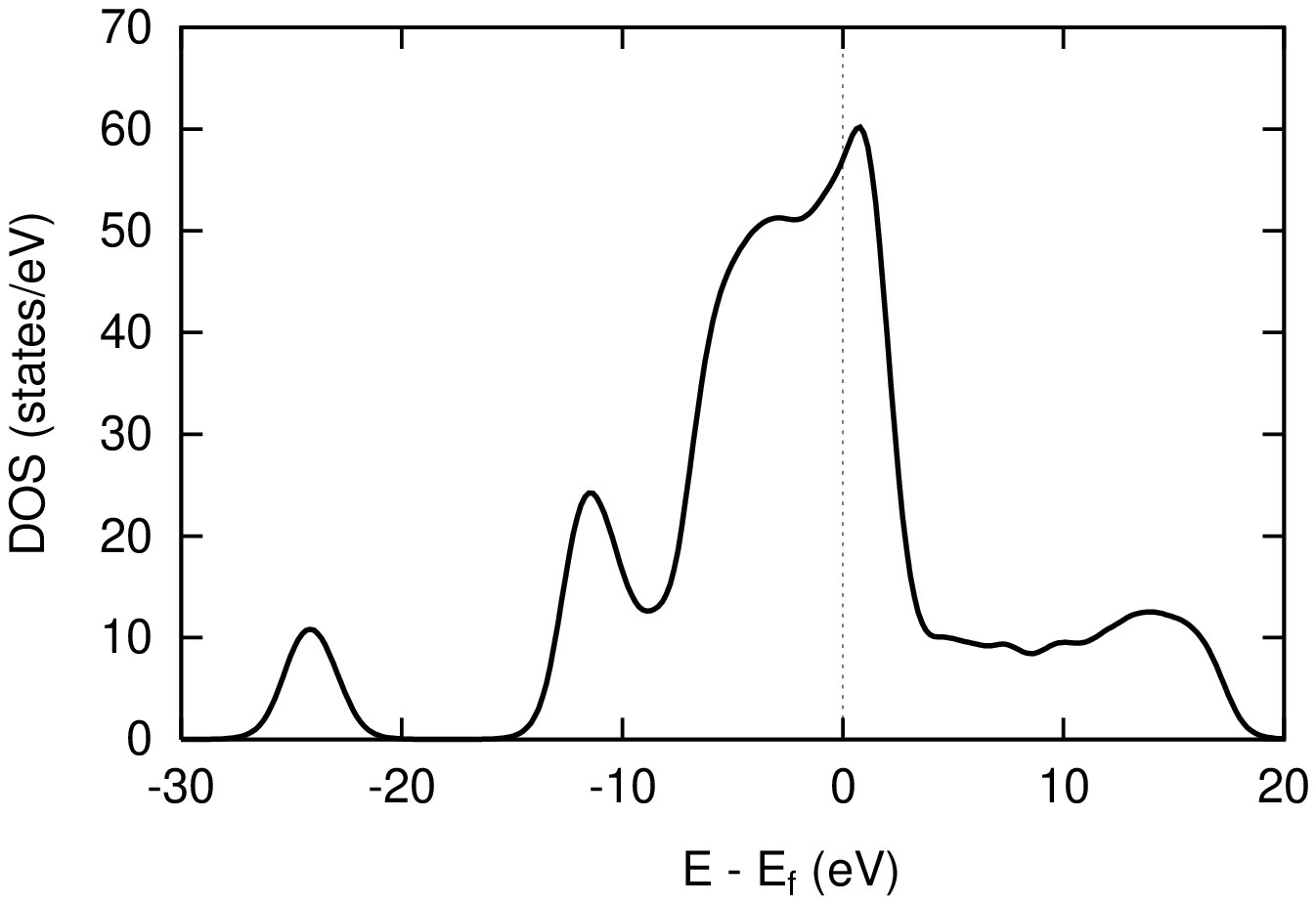,height=3.5in}}
\bigskip\centerline{\psfig{figure=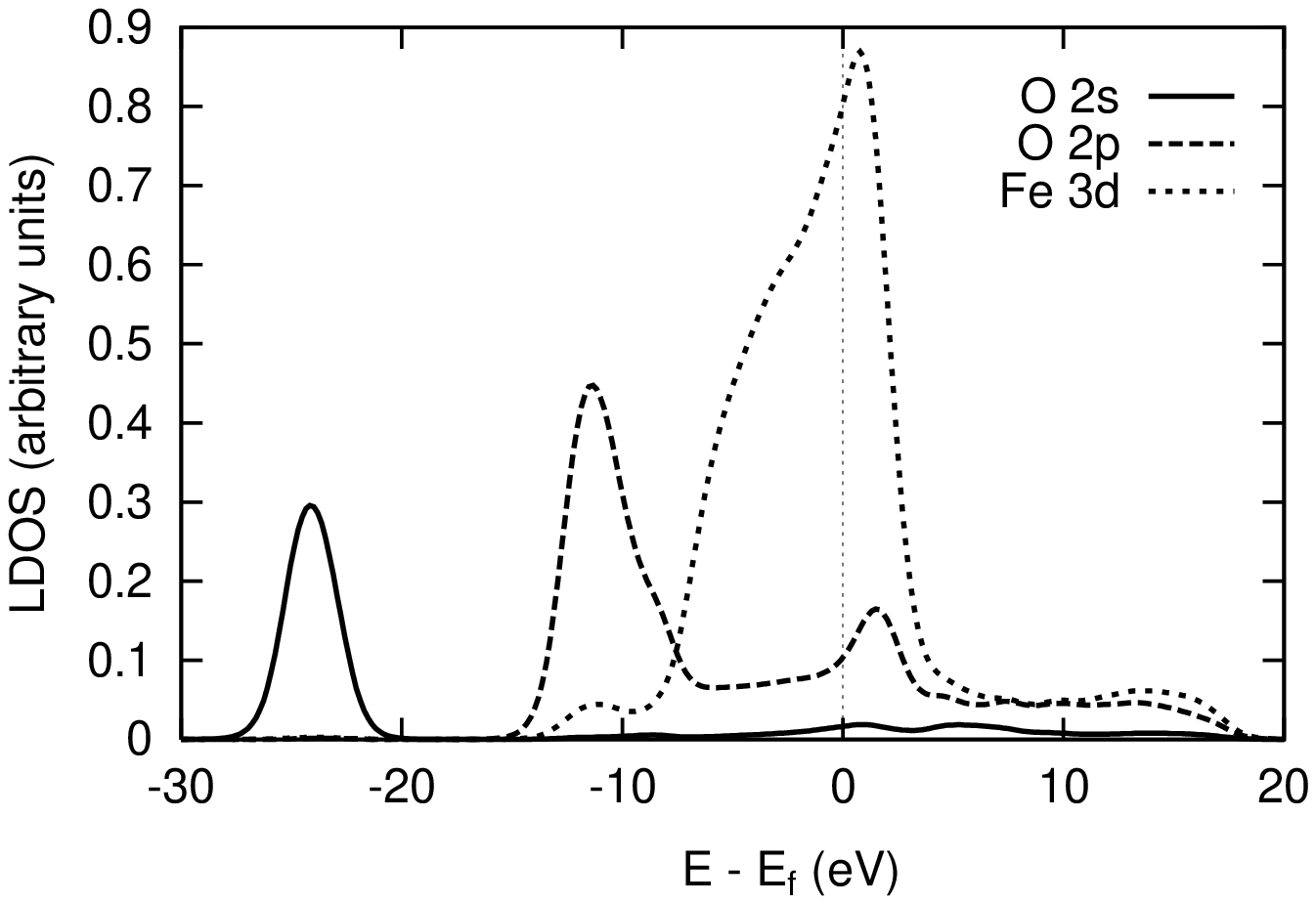,height=3.5in}}
 		
\newpage
\bigskip\centerline{FIGURE 12}
\bigskip\centerline{\psfig{figure=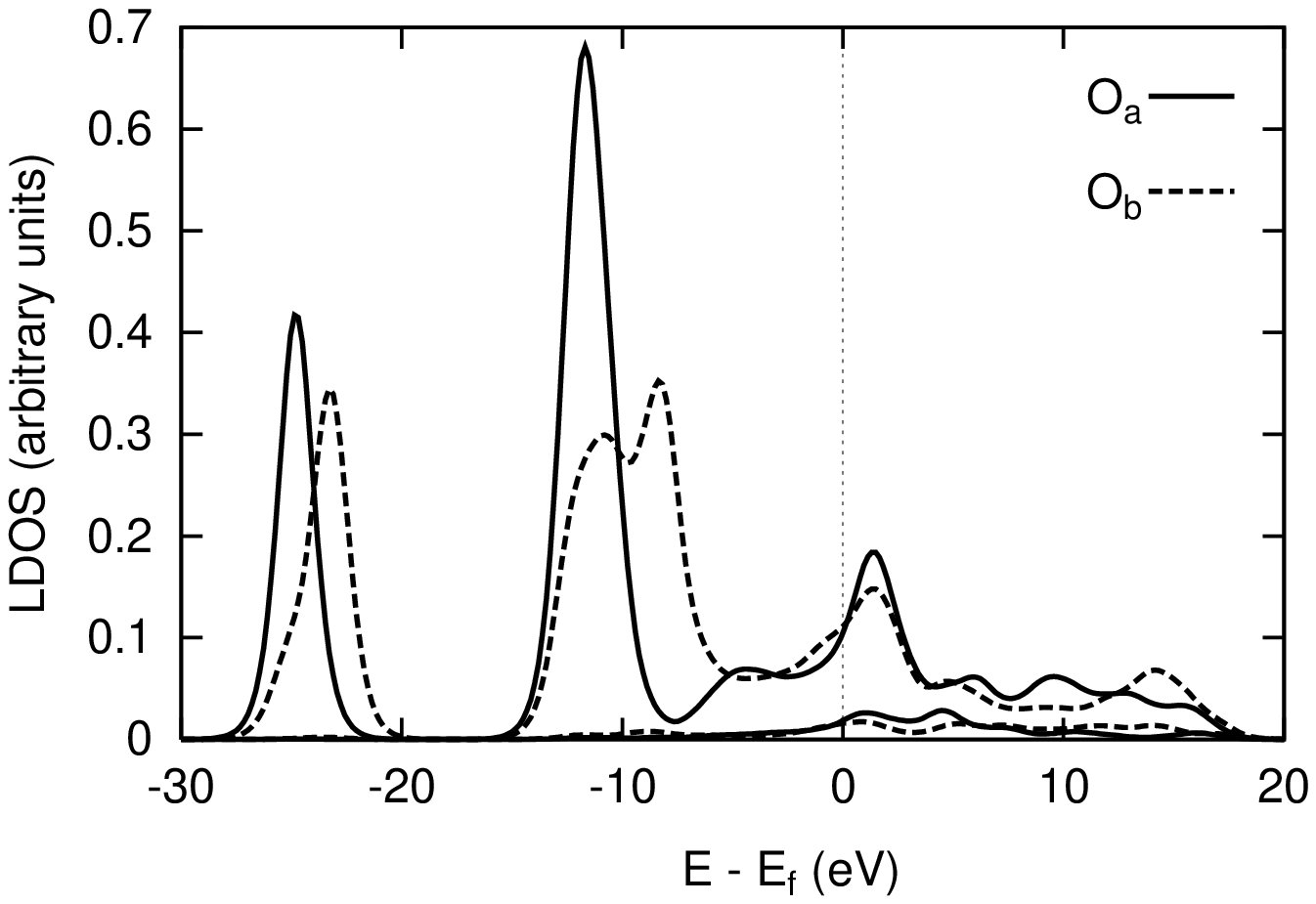,height=3.5in}}

\end{document}